\documentclass[12pt]{article}
\usepackage{amsmath}
\usepackage{amssymb}
\usepackage{graphicx}
\usepackage{ifthen}
\usepackage{verbatim}

\newlength{\captionwidth}
\newsavebox{\tempbox}
\newcommand{\mycaption}[2]{%
\par\vspace{10pt}\sbox{\tempbox}{Figure #1: #2}%
\ifthenelse{\lengthtest{\wd\tempbox>\captionwidth}}%
{\sbox{\tempbox}{Figure.#1:\ }%
\addtolength{\captionwidth}{-\wd\tempbox}%
\mbox{Figure #1:\ }\parbox[t]{\captionwidth}{\small\textit{#2}}}%
{Figure #1: {\small\textit{#2}}}}%

\numberwithin{equation}{section}

\newtheorem{proposition}{Proposition}

\allowdisplaybreaks
\begin{document}
\thispagestyle{empty}
%%%%%%%%%%%%%%%%%%%%%%
%%%%% TITLE PAGE %%%%%
%%%%%%%%%%%%%%%%%%%%%%
\begin{flushright}
\texttt{hep-th/0412327}\\
\texttt{OU-HET~512}\\
December 2004
\end{flushright}
\bigskip
\bigskip
%%%%% TITLE %%%%%
\begin{center}
{\Large \textbf{Five-Dimensional Supersymmetric Yang-Mills Theories}}
\end{center}
\begin{center}
{\Large \textbf{and Random Plane Partitions}}
\end{center}
\bigskip
\bigskip
%%%%% AUTHOR %%%%%
\renewcommand{\thefootnote}{\fnsymbol{footnote}}
\begin{center}
Takashi Maeda
\footnote{E-mail: \texttt{maeda@het.phys.sci.osaka-u.ac.jp}}$^1$, 
Toshio Nakatsu
\footnote{E-mail: \texttt{nakatsu@het.phys.sci.osaka-u.ac.jp}}$^1$, 
Kanehisa Takasaki
\footnote{E-mail: \texttt{takasaki@math.h.kyoto-u.ac.jp}}$^2$
and 
Takeshi Tamakoshi
\footnote{E-mail: \texttt{tamakoshi@het.phys.sci.osaka-u.ac.jp}}$^1$\\
\bigskip
{\small 
\textit{$^1$Department of Physics, Graduate School of Science, 
Osaka University,\\
Toyonaka, Osaka 560-0043, Japan\\
$^2$Graduate School of Human and Environmental Studies, 
Kyoto University, \\
Yoshida, Sakyo, Kyoto 606-8501, Japan
}}
\end{center}
\bigskip
\bigskip
\renewcommand{\thefootnote}{\arabic{footnote}}
%%%%% ABSTRACT %%%%%
\begin{abstract}
Five-dimensional $\mathcal{N}=1$ supersymmetric Yang-Mills theories 
are investigated from the viewpoint of random plane partitions. 
It is shown that random plane partitions are factorizable as 
$q$-deformed random partitions so that they admit the interpretations 
as five-dimensional Yang-Mills and as topological string amplitudes. 
In particular, they lead to the exact partition functions of 
five-dimensional $\mathcal{N}=1$ supersymmetric Yang-Mills 
with the Chern-Simons terms. 
We further show that some specific partitions, which we call 
the ground partitions, describe the perturbative regime of the 
gauge theories. We also argue their role in string theory. 
The gauge instantons give the deformation of the ground partition. 
\end{abstract}

\setcounter{footnote}{0}
\newpage
%%%%%%%%%%%%%%%%%%%%%%%%%%%%%%%%%%%%%%%%%%%%%%%%%%%%%%%%%%
%%%%%%%%%%%%%%%%%%%% TEXT %%%%%%%%%%%%%%%%%%%%%%%%%%%%%%%%
%%%%%%%%%%%%%%%%%%%%%%%%%%%%%%%%%%%%%%%%%%%%%%%%%%%%%%%%%%
\section{Introduction and summary}
\label{section1}

Recently it becomes possible 
\cite{Nekrasov, Nekrasov-Okounkov} to compute 
the exact partition functions of four-dimensional 
$\mathcal{N}=2$ supersymmetric gauge theories. 
Remarkably, the celebrated Seiberg-Witten solutions 
\cite{Seiberg-Witten} of the gauge theories  
emerge \cite{Nekrasov-Okounkov, Nakajima-Yoshioka} 
through the statistical models of random partitions.

$\mathcal{N}=2$ supersymmetric 
gauge theories are realized \cite{Geometric engineering}
by Type IIA strings on certain Calabi-Yau threefolds. 
The topological vertex \cite{topological vertex} 
provides a powerful method to compute 
all genus topological $A$-model partition functions 
on local toric Calabi-Yau threefolds. 
The string amplitudes on the relevant threefolds are 
computed \cite{Iqbal, Eguchi} by using this 
and shown to reproduce the gauge instanton contributions  
to the exact partition functions of the gauge theories. 
It turns out surprisingly that the topological vertex is 
identified \cite{Crystal} with the partition function of 
a model of random plane partitions \cite{Okounkov-Reshetikhin}. 
This interpretation is further explored in \cite{quantum foam}. 
The appearance of random plane partitions in string theory 
suggests their possible relation to the gauge theories.

Nevertheless, the exact solutions of the gauge theories 
have not been studied from the viewpoint of random plane partitions. 
In this article, we investigate 
five-dimensional $\mathcal{N}=1$ supersymmetric Yang-Mills theories 
from the perspective of random plane partitions. 
A plane partition $\pi=(\pi_{i j})_{i,j \geq 1}$ is  
an array of non-negative integers satisfying 
$\pi_{i j}\geq \pi_{i+1 j}$ and $\pi_{i j}\geq \pi_{i j+1}$, 
and identified with the three-dimensional Young diagram  
as depicted in Figure 1-(a). The diagram is also regarded 
as a sequence of partitions $\pi(m)$, where $m \in \mathbb{Z}$. 
See Figure 1-(b). Among the series of partitions, 
$\pi(0)$ will be called the main diagonal partition of $\pi$.  
We consider the following model 
of random plane partitions. 
\begin{eqnarray}
Z(q,Q)=\sum_{\pi}\,q^{|\pi|}\,Q^{|\pi(0)|}, 
\label{Z(q,Q)}
\end{eqnarray}
where $q$ and $Q$ are indeterminates. 
$|\pi|$ and $|\pi(0)|$ denote 
respectively the total numbers 
of cubes and boxes of the corresponding diagrams. 
By an identification of $q$ and $Q$ 
with the relevant string theory parameters,  
the partition function can be converted into 
topological $A$-model string amplitude 
on the local Calabi-Yau threefold 
$\mathcal{O}\oplus\mathcal{O}(-2)\to \mathbb{P}^1$. 
\begin{figure}[t]
\begin{center}
\includegraphics[scale=0.7]{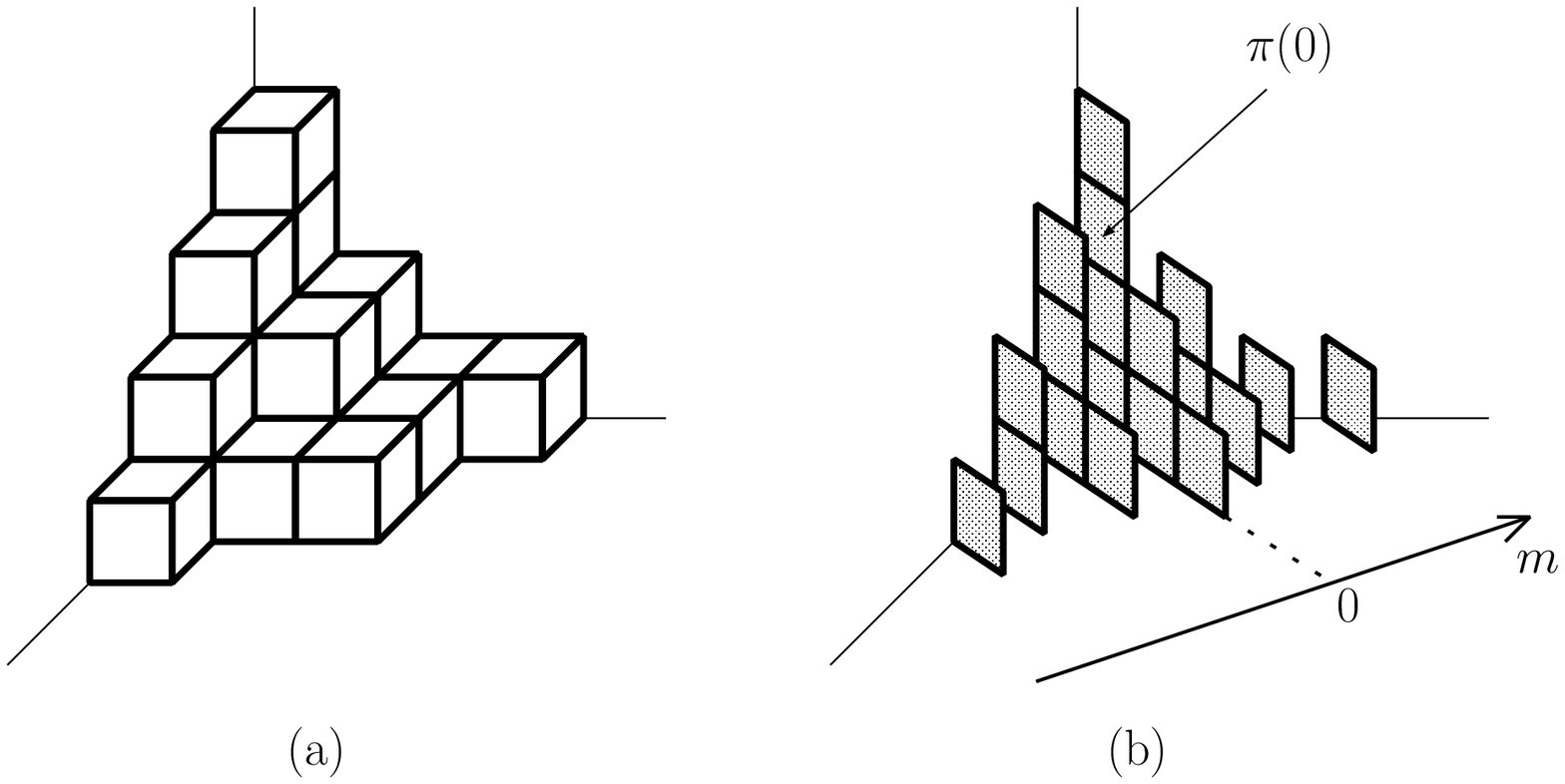}
\mycaption{1}{The three-dimensional Young diagram (a) 
and the corresponding sequence of partitions 
(the two-dimensional Young diagrams) (b).}
\end{center}\label{3d Young diagram}
\end{figure}

The above model has an interpretation 
as random partitions. It can be seen 
by rewriting the partition function as 
\begin{eqnarray}
Z(q,Q)=\sum_{\mu}\,Q^{|\mu|}\, 
\Bigl( \sum_{\pi(0)=\mu}\, q^{|\pi|} \Bigr), 
\label{Z(q,Q) random partitions}
\end{eqnarray}
where the summation over plane partitions in 
(\ref{Z(q,Q)}) are divided into two branches. 
The partitions $\mu$ are thought 
as the ensemble of the model by summing first over the 
plane partitions whose main diagonal partitions are $\mu$. 
As the model of random partitions, 
it becomes a $q$-deformation of the standard 
Plancherel measured random partitions.

It is well known that partitions are realized as states 
of two-dimensional free fermions. 
By using the folding \cite{Miwa-Jimbo} of a single complex fermion 
into $N$-component ones, each partition can be expressed 
as a set of $N$ charged partitions $(\lambda^{(r)},p_r)$, 
where $\lambda^{(r)}$ are partitions 
and $p_r$ are the $U(1)$ charges of the $N$-component fermions. 
This correspondence is used 
\cite{Nekrasov-Okounkov} to study the exact partition functions  
for four-dimensional $\mathcal{N}=2$ supersymmetric gauge theories. 
Among partitions, these coming from the charged empty partitions 
$(\lambda^{(r)},p_r)=(\emptyset,p_r)$ turn out to play special roles. 
We call them the ground partitions.  
Regarding the model as the $q$-deformed random partitions, 
we write the Boltzmann weights for the ground partitions 
by $Z_q^{SU(N)\,pert}(\{p_r\},Q)$. 
We will factor the partition function into  
\begin{eqnarray}
Z(q,Q)=
\sum_{\{p_r\}}\,
Z_q^{SU(N)\,pert}(\{p_r\},Q)\,
\sum_{\{\lambda^{(r)}\}} \,
Z_q^{SU(N)\,inst}(\{\lambda^{(r)}\},\{p_r\},Q).
\label{factored Z(q,Q)}
\end{eqnarray}

This factorization turns out useful to find out 
the gauge theoretical interpretations. 
The relevant field theory parameters are 
$a_r,\Lambda$ and $R$, where $a_r$ are the vacuum 
expectation values of the adjoint scalar in the vector 
multiplet and $\Lambda$ is the scale parameter of the 
underlying four-dimensional theory. $R$ is the radius of 
$S^1$ in the fifth dimension. 
We identify these parameters with 
$q,Q$ and $p_r$ in (\ref{factored Z(q,Q)}) as follows;
\begin{eqnarray}
q=e^{-\frac{2}{N}R\hbar},~~~~
Q=-(2R\Lambda)^2,~~~~
\widetilde{p}_r=a_r/\hbar,
\label{FT parameter introduction}
\end{eqnarray}
where $\widetilde{p}_r=p_r+\frac{1}{N}(r-\frac{N+1}{2})$. 
The parameter $\hbar$, 
which can be thought as a chemical potential 
for plane partitions, is often identified with the string 
coupling constant $g_{st}$.   
Via the above identifications we derive   
\begin{eqnarray}
Z_q^{SU(N)\,pert}(\{p_r\},Q)
\sum_{\{\lambda^{(r)}\}}
Z_q^{SU(N)\,inst}(\{\lambda^{(r)}\}, \{p_r\},Q)
=
Z^{SU(N)}_{\,5d \, \mbox{\tiny{SYM}}}
(\{a_r\};\,\Lambda,R,\hbar), 
\label{Z 5dSYM}
\end{eqnarray}
where the RHS is the exact partition function 
\cite{Nekrasov-Okounkov} for 
five-dimensional $\mathcal{N}=1$ supersymmetric 
$SU(N)$ Yang-Mills with the Chern-Simons term 
\cite{Chern-Simons}. 
The five-dimensional theory is living on 
$\mathbb{R}^4 \times S^1$. 
The Chern-Simons coupling constant $c_{cs}$ 
is quantized to $N$. The above can be said more 
precisely as follows; 
\begin{eqnarray}
Z_q^{SU(N)\,pert}(\{p_r\},\,Q)&=&
Z^{SU(N)\,pert}_{\,5d\,\mbox{\tiny{SYM}}}
(\{a_r\};\,\Lambda,R,\hbar),  
\label{Z 5dSYM pert} \\[1.5mm]
\sum_{\{\lambda^{(r)}\}}
Z_q^{SU(N)\,inst}
(\{\lambda^{(r)}\},\{p_r\},\,Q)&=&
Z^{SU(N)\,inst}_{\,5d\,\mbox{\tiny{SYM}}}
(\{a_r\};\,\Lambda,R,\hbar), 
\label{Z 5dSYM inst}
\end{eqnarray}
where the RHSs are respectively the perturbative part 
and the instanton part of the exact partition function.

The gauge theory partition function is obtained 
as the $\hbar \rightarrow 0$ limit of the above partition function. 
The ground partition in (\ref{Z 5dSYM}) becomes very large 
since its size is $\sim p_r^2$ and 
the $U(1)$ charges $p_r$ scale as $\hbar^{-1}$ 
at the field theory limit. 
The equality (\ref{Z 5dSYM pert}) shows that these ground 
partitions describe the perturbative regime of the Coloumb branch.  
$\lambda^{(r)}$ in (\ref{Z 5dSYM}) represent   
$U(1)$ instantons with $c_2=|\lambda^{(r)}|$ 
on a non-commutative $\mathbb{R}^4$ \cite{Nakajima}. 
They can be viewed as excitations from the ground partition. 
These excitations emerge as the quantum (non-perturbative) 
deformation of the perturbative Coloumb branch 
when $\lambda^{(r)}$ are comparable to the ground partition.

We start Section 2 with a brief review on 
the transfer matrix approach \cite{Okounkov-Reshetikhin} 
to random plane partitions. 
The perspective of the $q$-deformation is emphasized. 
In Section 3 we examine 
the factorization (\ref{factored Z(q,Q)}) 
for the case of $SU(2)$ and confirm 
the equality (\ref{Z 5dSYM inst}).  
Section 4 is devoted to prove the equality 
(\ref{Z 5dSYM pert}) for this case. 
The discussion is also applicable to the cases of the higher 
rank gauge groups. 
In Section 5 we treat 
the cases of $SU(N)$. 
The instanton parts can be converted 
to topological $A$-model string amplitudes on the relevant  
local Calabi-Yau threefolds. These amplitudes are summarized 
in Appendix.

%%%%%%%%%%%%%%%%%%%%%%%%%%%%%%%%%%%%%%%%%%%%%%%%%%%%%%%%%
%%%%%%%%%%%%%%%%%%%%%%%%%%%%%%%%%%%%%%%%%%%%%%%%%%%%%%%%%
\section{Random plane partitions and $U(1)$ gauge theory}
\label{section2}

A plane partition $\pi$ is an array of non-negative integers 
\begin{eqnarray}
\begin{array}{cccc}
\pi_{11} & \pi_{12} & \pi_{13} & \cdots \\
\pi_{21} & \pi_{22} & \pi_{23} & \cdots \\
\pi_{31} & \pi_{32} & \pi_{33} & \cdots \\
\vdots & \vdots & \vdots & ~
\end{array}
\label{pi}
\end{eqnarray}
satisfying 
$\pi_{ij}\geq \pi_{i+1 j}$ and 
$\pi_{ij}\geq \pi_{i j+1}$ for 
all $i,j \geq 1$. 
Plane partitions are identified 
with the three-dimensional 
Young diagrams. The three-dimensional diagram $\pi$ 
is a set of unit cubes such that $\pi_{ij}$ cubes 
are stacked vertically on each $(i,j)$-element of $\pi$. 
The size of $\pi$ is $|\pi| \equiv \sum_{i,j \geq 1}\pi_{ij}$, 
which is the total number of cubes of the diagram. 
Each diagonal slice of $\pi$ becomes a partition, 
that is, a sequence of weakly decreasing non-negative integers.  
Let $\pi(m)$ be a partition along the $m$-th diagonal slice. 
\begin{eqnarray}
\pi(m)= 
\left\{
\begin{array}{ll}
(\pi_{1\,\,m+1},\pi_{2\,\, m+2},\pi_{3\,\, m+3},\cdots) & 
~~\mbox{for}~m \geq 0 \\
(\pi_{-m+1\,\, 1},\pi_{-m+2\,\, 2},\pi_{-m+3\,\, 3},\cdots) & 
~~\mbox{for}~m \leq -1.
\end{array}
\right.
\end{eqnarray}
In particular 
$\pi(0)=(\pi_{11}\geq\pi_{22}\geq\pi_{33}\geq\cdots\geq 0)$ 
is the main diagonal partition. 
The series of partitions $\pi(m)$ satisfies the condition
\begin{eqnarray}
\cdots \prec \pi(-2) \prec \pi(-1) \prec 
\pi(0) \succ \pi(1) \succ \pi(2) \succ \cdots,
\end{eqnarray}
where $\mu \succ \nu$ means the following interlace relation 
between two partitions 
$\mu=(\mu_1 \geq \mu_2 \geq \cdots \geq 0)$ 
and 
$\nu=(\nu_1 \geq \nu_2 \geq \cdots \geq 0)$ 
\begin{eqnarray}
\mu \succ \nu ~~~
\Longleftrightarrow ~~~
\mu_1 \geq \nu_1 \geq \mu_2 \geq \nu_2 
\geq \mu_3 \geq \cdots.
\end{eqnarray} 
We have $|\pi|=\sum_{m=-\infty}^{+\infty}|\pi(m)|$,  
where the size of a partition $\mu$ is also denoted by  
$|\mu| \equiv \sum_{i \geq 1}\mu_i$.

%%%%%%%%%%%%%%%%%%%%%%%%%%%%%%%%%%%%%%%%%%%%%%%%%%%%%%%%%
\subsection{Random plane partitions}

A model of random plane partitions 
relevant to describe five-dimensional ${\cal N}=1$ 
supersymmetric $U(1)$ gauge theory is defined 
by the following partition function.
\begin{eqnarray}
Z_q^{U(1)}(Q)
&\equiv& 
\sum_{\pi}\, q^{|\pi|}Q^{|\pi(0)|}. 
\label{Z U(1)}
\end{eqnarray}
The Boltzmann weight consists of two parts. 
The first contribution comes from the energy 
of plane partitions $\pi$, and 
the second contribution is a chemical potential 
for the main diagonal partitions $\pi(0)$.  
To contact with the $U(1)$ gauge theory 
we will identify the indeterminates $q$ and $Q$ 
with the following field theory parameters. 
\begin{eqnarray}
q=e^{-2R\hbar},~~~
Q=-(2R\Lambda)^2, 
\label{U(1) q and Q}
\end{eqnarray}
where $R$ is the radius of $S^1$ in the fifth dimension 
and $\Lambda$ denotes the scale parameter of the 
underlying four-dimensional theory. The parameter 
$\hbar$ is often identified with 
string coupling constant $g_{st}$.

%%%%%%%%%%%%%%%%%%%%%%%%%%%%%%%%%%%%%%%%%%%%%%%%%%%%%%%%%%%%
\subsubsection{Transfer matrix approach}

The transfer matrix approach 
\cite{Okounkov-Reshetikhin} 
allows us to express the random plane partitions 
(\ref{Z U(1)}) in terms of two-dimensional conformal 
field theory (2$d$ free fermion system).  
The partition function can be computed exactly 
by using the standard technique of 2$d$ CFT.

It is well known that partitions 
are realized as states of 2$d$ free fermions 
by using the Maya diagrams. 
Let 
$\psi(z)=\sum_{k \in \mathbb{Z}+\frac{1}{2}}
\psi_k z^{-k-\frac{1}{2}}$
and 
$\psi^*(z)=\sum_{k \in \mathbb{Z}+\frac{1}{2}}
\psi^*_k z^{-k-\frac{1}{2}}$
be complex fermions with the anti-commutation relations 
\begin{eqnarray}  
\left\{ \psi_k,\psi^*_l \right\}
=\delta_{k+l,0},~~~~
\left\{ \psi_k,\psi_l \right\}
=\left\{ \psi_k^*,\psi_l^* \right\}
=0. 
\end{eqnarray}
Let 
$\mu=(\mu_1, \mu_2, \cdots)$ be a partition. 
The Maya diagram is the set 
$S_{\mu}$ which is defined by 
\begin{eqnarray}
S_{\mu}\equiv \left\{
x_{i}(\mu) 
= \mu_i-i+\frac{1}{2} ~;  
i \geq 1 
\right\} 
~~~~~
\left(
\subset \mathbb{Z}+\frac{1}{2}
\right).
\label{Maya diagram S}
\end{eqnarray}
The correspondence with the Young diagram 
is depicted in Figure 2. 
\begin{figure}[t]
\begin{center}
\includegraphics[scale=0.5]{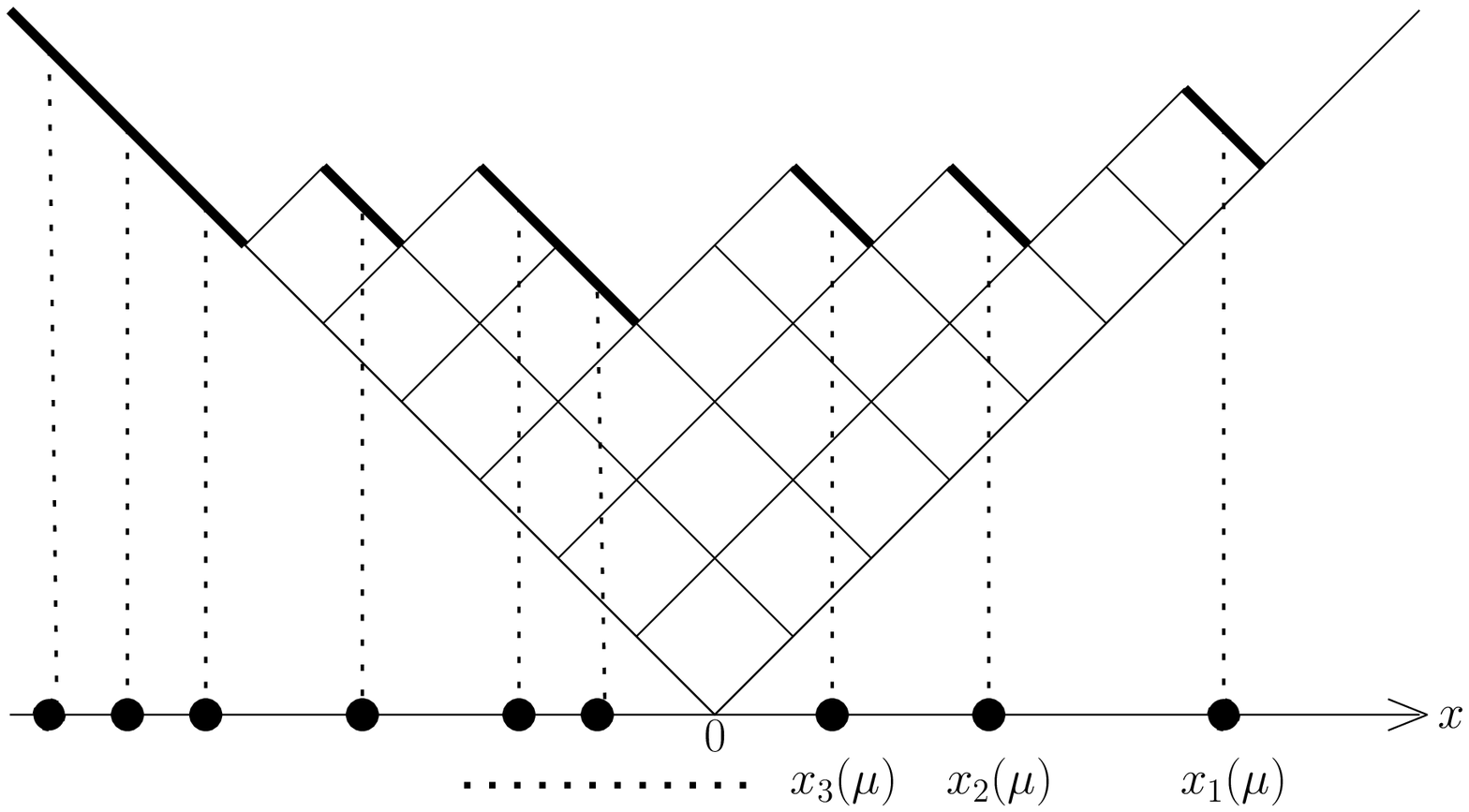}
\mycaption{2}{The correspondence between 
the Maya diagram and 
the Young diagram of $\mu=(7,5,4,2,2,1)$.
Elements of the Maya diagram 
are denoted by $\bullet$.} 
\end{center}\label{Maya diagram}
\end{figure}
By using the Maya diagram 
the partition can be mapped to 
the following fermion state. 
\begin{eqnarray}
|\mu;n \rangle 
=
\psi_{-x_1(\mu)-n}\psi_{-x_2(\mu)-n}
\cdots \psi_{-x_{l(\mu)}(\mu)-n}
\psi^*_{-l(\mu)+\frac{1}{2}+n}
\psi^*_{-l(\mu)+\frac{3}{2}+n}
\cdots \psi^*_{-\frac{1}{2}+n} 
|\emptyset ;n \rangle, 
\label{|mu;n>}
\end{eqnarray}
where $l(\mu)$ is the length of $\mu$, that is, 
the number of the non-zero $\mu_i$. 
In (\ref{|mu;n>}) the state $|\emptyset ;n\rangle$ is 
the ground state of the charge $n$ sector. 
It is defined by the conditions  
\begin{eqnarray}
&&
\psi_{k}|\emptyset ;n \rangle=0~~~\mbox{for}~k > -n,
\nonumber \\
&&
\psi_{k}^*|\emptyset ;n \rangle=0~~~\mbox{for}~k > n.
\label{|phi;n>}
\end{eqnarray}
We mainly consider the $n=0$ sector in the below.

The basic ingredient of the transfer matrix approach 
is the following evolution operator at a discretized time 
$m \in \mathbb{Z}$.
\begin{eqnarray}
\Gamma(m)
&\equiv& 
\left\{
\begin{array}{cl}
\exp\Bigl(
\displaystyle{
\sum_{k=1}^{+\infty}
\frac{1}{k}q^{k(m+\frac{1}{2})}J_{-k}}
\Bigr)
& 
~~~\mbox{for}~m \geq 0 \\[1.5mm]
\exp\Bigl(
\displaystyle{
\sum_{k=1}^{+\infty}
\frac{1}{k}q^{-k(m+\frac{1}{2})}J_{k}}
\Bigr)
& 
~~~\mbox{for}~m \leq -1,  
\end{array}
\right.
\label{gamma m}
\end{eqnarray}
where $J_{\pm k}$ are the modes of the standard $U(1)$ current 
\begin{eqnarray}
:\psi \psi^*:(z)=\sum_{n}J_nz^{-n-1}.
\end{eqnarray}
Implications of the above operators 
in random plane partitions can be 
understood from their matrix elements: 
For $m \geq 0$, 
\begin{eqnarray}
\langle \mu;0|
\Gamma(m)
|\nu;0 \rangle 
=
\left\{
\begin{array}{cl}
q^{(m+\frac{1}{2})(|\mu|-|\nu|)}  &
\mu \succ \nu    \\
0 & 
\mbox{otherwise}, 
\end{array}
\right.
\label{gamma m on mu 1}
\end{eqnarray}
and for $m \leq -1$,  
\begin{eqnarray}
\langle \mu;0|
\Gamma(m)
|\nu;0 \rangle 
=
\left\{
\begin{array}{cl}
q^{(m+\frac{1}{2})(|\mu|-|\nu|)}  &
\mu \prec \nu    \\
0 & 
\mbox{otherwise}. 
\end{array}
\right.
\label{gamma m on mu 2}
\end{eqnarray}
It follows from (\ref{gamma m on mu 1}) 
and (\ref{gamma m on mu 2}) that the partition 
function is expressed as 
\begin{eqnarray}
Z_q^{U(1)}(Q)&=&
\langle \emptyset ;0|
\Bigl\{ 
\prod_{m \leq -1} \Gamma(m)
\Bigr\}\,
Q^{L_0}\,
\Bigl\{
\prod_{m \geq 0}  \Gamma(m)
\Bigr\}
| \emptyset ; 0\rangle.
\label{Z U(1) transfer matrix}
\end{eqnarray}
This expression allows us to compute 
the partition function 
by using the standard technique of 
2$d$ CFT. It becomes 
\begin{eqnarray}
Z_q^{U(1)}(Q)&=&\prod_{n=1}^{+\infty}
\,\frac{1}{(1-Qq^n)^{n}}.
\label{MacMahon}
\end{eqnarray}

It is also possible to regard $\hbar$ 
as a chemical potential for plane partitions.  
Thermodynamic limit of (\ref{Z U(1)}) is obtained 
by letting $\hbar \rightarrow 0$. 
Let us recall that the mean values 
of $|\pi|$ and $|\pi(0)|$ 
are respectively given by 
$q\frac{\partial}{\partial q}\ln Z_q^{U(1)}$ 
and $Q\frac{\partial}{\partial Q}\ln Z_q^{U(1)}$. 
It follows from (\ref{MacMahon}) 
that they behave 
$\left \langle |\pi| \right \rangle =O(\hbar^{-3})$ 
and 
$\left \langle |\pi(0)| \right \rangle =O(\hbar^{-2})$ 
as $\hbar \rightarrow 0$ ($q \rightarrow 1$). 
Therefore a typical plane partition $\pi$ 
at the limit $\hbar \rightarrow 0$ is a plane partition 
of order $\hbar^{-3}$, 
and its main diagonal partition $\pi(0)$ 
becomes a partition of order $\hbar^{-2}$.

%%%%%%%%%%%%%%%%%%%%%%%%%%%%%%%%%%%%%%%%%%%%%%%%%%%%%%%%%%%
\subsubsection{$q$-deformed random partitions}

We can interpret the random plane partitions 
(\ref{Z U(1)}) as a model of random partitions. 
It is identified with a $q$-deformation of 
the standard Plancherel measured random partitions. 
(The deformation is different from \cite{Nekrasov-Okounkov}.) 
To see this, 
we rewrite the partition function 
by using the Schur functions. 
An insertion of the unity 
$1=\sum_{\mu}|\mu;0 \rangle \langle \mu ;0|$ 
factorizes (\ref{Z U(1) transfer matrix}) into 
\begin{eqnarray}
Z_q^{U(1)}(Q)&=&
\sum_{\mu}
Q^{|\mu|}
\langle \emptyset ;0|
\prod_{m \leq -1}
\Gamma(m)
|\mu; 0\rangle 
\langle \mu;0| 
\prod_{m \geq 0}
\Gamma(m)
|\emptyset ;0\rangle.  
\label{factorized Z U(1) transfer matrix}
\end{eqnarray}
The matrix elements in the above turn to be 
\begin{eqnarray}
\langle \emptyset ;0|
\prod_{m \leq -1}\Gamma(m) 
| \mu; 0 \rangle 
&=&
\langle \emptyset ;0 | 
\prod_{k=1}^{+\infty}
\exp \Bigl( 
\frac{1}{k}
\sum_{i=1}^{+\infty}
q^{k(i-\frac{1}{2})}
J_k \Bigr) 
| \mu; 0 \rangle 
\nonumber \\
&=&
s_{\mu}(q^{\frac{1}{2}},q^{\frac{3}{2}},\cdots), 
\label{schur mu 1} \\
\langle \mu;0|
\prod_{m \geq 0}\Gamma(m) 
| \emptyset; 0 \rangle 
&=&
\langle \mu;0 | 
\prod_{k=1}^{+\infty}
\exp \Bigl( 
\frac{1}{k}
\sum_{i=1}^{+\infty}
q^{k(i-\frac{1}{2})}
J_{-k} \Bigr) 
| \emptyset; 0 \rangle 
\nonumber \\
&=&
s_{\mu}(q^{\frac{1}{2}},q^{\frac{3}{2}},\cdots), 
\label{schur mu 2}
\end{eqnarray} 
where 
$s_{\mu}(q^{\frac{1}{2}},q^{\frac{3}{2}},\cdots)$ 
is the Schur function 
$s_{\mu}(x_1,x_2,\cdots)$ specialized at 
$x_i=q^{i-\frac{1}{2}}$ ($i \geq 1$). 
Therefore we obtain 
\begin{eqnarray}
Z_q^{U(1)}(Q)
&=&
\sum_{\mu}Q^{|\mu|}
s_{\mu}(q^{-\rho})^2, 
\label{Z U(1) schur}
\end{eqnarray}
where the multiple index 
$\rho \equiv (-\frac{1}{2},-\frac{3}{2}, 
\cdots, -i+\frac{1}{2},\cdots)$ is used. 
The expression (\ref{Z U(1) schur}) allows us 
to interpret (\ref{Z U(1)}) 
as a model of $q$-deformed random partitions. 
It is also clear from  
(\ref{factorized Z U(1) transfer matrix}) 
that partitions $\mu$ are the main diagonal 
partitions $\pi(0)$.

The four-dimensional limit of the model  
is obtained by letting $R \rightarrow 0$ 
under the identification (\ref{U(1) q and Q}). 
To obtain the limit of (\ref{Z U(1) schur}) 
the following product formula of the Schur function 
\cite{Macdonald} becomes useful. 
\begin{eqnarray}
s_{\mu}(q^{-\rho})
&=&
q^{-\frac{1}{4}\kappa(\mu)}
 \prod_{(i,j)\in \mu}\frac{1}{q^{-\frac{h(i,j)}{2}}
-q^{\frac{h(i,j)}{2}}}\,, 
\label{hook length formula 2}
\end{eqnarray}
where $h(i,j)$ 
is the hook length of the box $(i,j)$ 
in the Young diagram, 
and 
$\kappa(\mu) \equiv 2\sum_{(i,j)\in \mu}(j-i)$. 
By using (\ref{hook length formula 2}) 
we can see from (\ref{Z U(1) schur}) that 
\begin{eqnarray}
\lim_{R \rightarrow 0}\,Z_q^{U(1)}(Q)
&=&
\sum_{\mu} 
\left( \frac{\Lambda}{\hbar}\right)^{2|\mu|}
\frac{(-1)^{|\mu|}}{(\prod_{(i,j)\in\mu}h(i,j))^2}\,. 
\end{eqnarray}
This is the random partitions \cite{Nekrasov-Okounkov,Nakajima-Yoshioka}
which describes four-dimensional 
${\cal N}=2$ supersymmetric $U(1)$ gauge theory 
on non-commutative $\mathbb{R}^4$.

%%%%%%%%%%%%%%%%%%%%%%%%%%%%%%%%%%%%%%%%%%%%%%%%%%
\subsubsection{Interpretation as topological string amplitude}

The partition function (\ref{Z U(1)}) can be converted 
to topological string amplitude on a certain 
non-compact Calabi-Yau threefold by identifying $q$ and $Q$ 
with the string theory parameters. 
Let us consider the topological $A$-model on 
$\mathcal{O}\oplus\mathcal{O}(-2)\to \mathbb{P}^1$.  
All genus $A$-model partition function on this local geometry 
is computed \cite{Gopakumar} from the $M$-theory viewpoint.
It is given by 
\begin{eqnarray}
Z_{string}^{U(1)}(q,Q)
&=&
\prod_{n=1}^{\infty}\frac{1}{(1-Qq^n)^n},  
\label{U(1) string}
\end{eqnarray}
where the string coupling constant $g_{st}$ and 
the K\"ahler parameter $t$ of the $\mathbb{P}^1$  
appear as $q=e^{-g_{st}}$ and $Q=e^{-t}$.

The topological vertex \cite{topological vertex} 
makes it possible to compute all genus $A$-model 
partition functions on local toric Calabi-Yau threefolds 
by using diagrammatic techniques like the Feynman rules, 
where the diagrams are the dual toric diagrams 
of the local geometries. The diagram for the above 
local geometry is described in Figure 3. 
A computation using the topological vertices presents  
the amplitude in the same form as (\ref{Z U(1) schur}). 
In that expression, partitions $\mu$ are 
attached to the $\mathbb{P}^1$. 
\begin{figure}[htb]
\begin{center}
\includegraphics[scale=0.7]{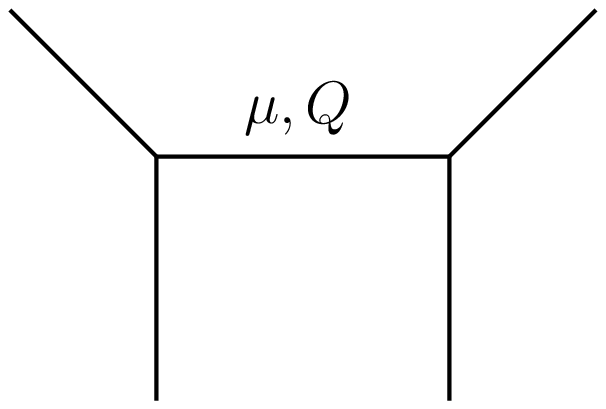}
\mycaption{3}{The diagram for 
$\mathcal{O}\oplus\mathcal{O}(-2)\to \mathbb{P}^1$.
$Q=e^{-t}$, where $t$ is the K\"ahler parameter 
of the $\mathbb{P}^1$. Partitions $\mu$ are 
attached to the $\mathbb{P}^1$.}
\end{center}\label{U(1) geometry}
\end{figure}
%
%

%%%%%%%%%%%%%%%%%%%%%%%%%%%%%%%%%%%%%%%%%%%%%%%%%%%%%%%%%%%%%
%%%%%%%%%%%%%%%%%%%%%%%%%%%%%%%%%%%%%%%%%%%%%%%%%%%%%%%%%%%%%
\section{$SU(2)$ Yang-Mills and random plane partitions}
\label{section3}

%%%%%%%%%%%%%%%%%%%%%%%%%%%%%%%%%%%%%%%%%%%%%%%%%%%%%%%%%%%%
\subsection{Multi-component fermion}

The Fock representation of a single complex fermion 
has \cite{Miwa-Jimbo} an alternative realization 
by exploiting $N$ complex fermions: 
\begin{eqnarray}
\psi_{k}^{(r)},\ \psi_{l}^{(s)*},
\ \ k,l\in\mathbb{Z}+\frac{1}{2},\ r,s=1,2,\ldots,N, 
\end{eqnarray}
with the anti-commutation relations
\begin{eqnarray}
\{\psi_{k}^{(r)},\psi_{l}^{(s)*}\} 
= \delta_{r,s}\delta_{k+l,0}\ .
\end{eqnarray}
$\psi^{(r)}(z)$ and $\psi^{(s)*}(z)$ 
are obtained from $\psi(z)$ and $\psi^*(z)$ 
by the following identifications. 
\begin{eqnarray}
\psi_{k}^{(r)} = \psi_{N(k-\xi_{r})},~~~ 
\psi_{l}^{(s)*} = \psi_{N(l+\xi_{s})},
\label{N-1}
\end{eqnarray}
where 
$\xi_r 
\equiv 
\frac{1}{N}\left(r-\frac{N+1}{2}\right)$.

It is convenient to consider partitions paired 
with the $U(1)$ charges. 
We denote such a charged partition by $(\mu,n)$, 
where $\mu$ is a partition and $n$ is the $U(1)$ charge. 
The states $|\mu;\,n \rangle$ constitute 
bases of the Fock space of a single complex fermion.  
Thanks to the above realization of $N$ fermions, 
we can express $(\mu,n)$ uniquely by means of 
$N$ charged partitions $(\lambda^{(r)},p_r)$ and vice versa 
through the identification  
\begin{eqnarray}
|\mu;\,n\rangle &=& 
\bigotimes_{r=1}^{N} |\lambda^{(r)};\,p_r\rangle_{(r)}. 
\label{N-1 states}
\end{eqnarray}
The conservation of the $U(1)$ charges implies 
$\sum_{r=1}^Np_r=n$. 
The correspondence between the partitions 
$\mu$ and $\lambda^{(r)}$ can be described explicitly 
by using the Maya diagrams (shifted by the $U(1)$ charges). 
It becomes  
\begin{eqnarray}
\Bigl \{
x_i(\mu)+n~; i\geq 1
\Bigr \}
&=& 
\bigcup_{r=1}^N~
\Bigl \{
N(x_{i_r}(\lambda^{(r)})+\widetilde{p}_r)~ 
;i_r \geq 1 
\Bigr \}, 
\label{N-1 Maya diagrams}
\end{eqnarray}
where  $\widetilde{p}_r \equiv p_r+\xi_r$.
In particular, when $n=0$, 
the following information on the partitions 
is obtainable from the above correspondence 
by applying the method of power-sums 
\cite{Nekrasov-Okounkov}.
\begin{eqnarray}
|\mu| &=& 
N\sum_{r=1}^{N}
|\lambda^{(r)}|
+\frac{N}{2}\sum_{r=1}^{N}p_{r}^2
+\sum_{r=1}^{N}rp_r, 
\nonumber 
\\
\kappa(\mu) &=& 
N^2\sum_{r=1}^{N}\kappa(\lambda^{(r)})
+2N^2\sum_{r=1}^{N}\widetilde{p}_r |\lambda^{(r)}|
+\frac{N^2}{3}\sum_{r=1}^{N}\widetilde{p}_r^3. 
\label{N-1 size k}
\end{eqnarray}

The Maya diagram $S_{\lambda^{(r)}}$ is considered 
as the subset of $S_{\mu}$ by (\ref{N-1 Maya diagrams}). 
It is also a subset of $N \mathbb{Z}+r-\frac{1}{2}$
($\subset \mathbb{Z}+\frac{1}{2}$).
We may colour the set $N \mathbb{Z}+r-\frac{1}{2}$ 
just by attaching the number $r$ 
to all the elements. 
$S_{\lambda^{(r)}}$ gets coloured by $r$. 
(\ref{N-1 Maya diagrams}) shows that 
$S_{\mu}$ has $N$ colours, where the number $r$ 
is attached to these elements coming from 
$S_{\lambda^{(r)}}$.

%%%%%%%%%%%%%%%%%%%%%%%%%%%%%%%%%%%%%%%%%%%%%%%%%%%
%%%%%%%%%%%%%%%%%%%%%%%%%%%%%%%%%%%%%%%%%%%%%%%%%%%
\subsection{$SU(2)$ Yang-Mills from random plane partitions}

Let us begin with the case of two-component fermions. 
Owing to the identification (\ref{N-1 states}) 
any charged partition $(\mu,n)$ can be expressed by 
$(\lambda^{(1)},p_1)$ and $(\lambda^{(2)},p_2)$. 
We consider the neutral case, that is, $n=0$. 
$(\lambda^{(1)},p)$ and $(\lambda^{(2)},-p)$   
determine a partition in the neutral sector. 
We denote such a partition by 
$\mu(\lambda^{(1)},\lambda^{(2)};\,p)$. The corresponding 
Maya diagram $S_{\mu(\lambda^{(1)},\lambda^{(2)};\,p)}$ 
can be read from (\ref{N-1 Maya diagrams}) as 
\begin{eqnarray}
&&
S_{\mu(\lambda^{(1)},\lambda^{(2)};\,p)}
\nonumber \\
&&
~~=
\left\{2\left(x_{i_1}(\lambda^{(1)})+p-\frac{1}{4}\right)
~;i_1 \geq 1\right\}
\cup 
\left\{2\left(x_{i_2}(\lambda^{(2)})-p+\frac{1}{4}\right)
~;i_2 \geq 1\right\}. 
\label{2-1 Maya diagrams}
\end{eqnarray}
In the present case,  
the quantities (\ref{N-1 size k}) turn to be 
\begin{eqnarray}
|\mu(\lambda^{(1)},\lambda^{(2)};p)|
&=& 
2(|\lambda^{(1)}|+|\lambda^{(2)}|)+p(2p-1),
\notag\\[2mm]
\kappa(\mu(\lambda^{(1)},\lambda^{(2)};p))&=&
4(\kappa(\lambda^{(1)})
+\kappa(\lambda^{(2)}))
+2(4p-1)(|\lambda^{(1)}|-|\lambda^{(2)}|). 
\label{2-1 size k}
\end{eqnarray}

We regard the random plane partitions (\ref{Z U(1)}) 
as the $q$-deformed random partitions 
via the expression (\ref{Z U(1) schur}) 
and rewrite the partition function as 
\begin{eqnarray}
Z_q^{U(1)}(Q) &=&
\sum_{p \in \mathbb{Z}} 
\sum_{\lambda^{(1)},\lambda^{(2)}} 
Q^{\left|\mu(\lambda^{(1)},\lambda^{(2)};\,p)\right|}
s_{\mu(\lambda^{(1)},\lambda^{(2)};\,p)}(q^{-\rho})^2.  
\label{Z U(1) two-component}
\end{eqnarray}
Subsequently we will further factor the above 
into the following form. 
\begin{eqnarray}
Z_q^{U(1)}(Q)
&=& 
\sum_{p\in\mathbb{Z}}
Z_q^{pert}(p,Q)
\sum_{\lambda^{(1)},\lambda^{(2)}}
Z_q^{inst}(\lambda^{(1)},\lambda^{(2)},p,\,Q).  
\label{Z SU(2) factorization pre}
\end{eqnarray}

%%%%%%%%%%%%%%%%%%%%%%%%%%%%%%%%%%%%%%%%%%%%%%%%%%
\subsubsection{$SU(2)$ ground partitions}

We first consider the cases consisting 
of the empty partitions    
\begin{eqnarray}
\lambda^{(1)}=\lambda^{(2)}=\emptyset.
\end{eqnarray}
For each $p \in \mathbb{Z}$, the corresponding partition 
$\mu(\emptyset,\emptyset;\,p)$ is named 
the $SU(2)$ ground partition.  
We will show in the next section that 
these partitions are responsible 
to the perturbative gauge theory. 
It follows from (\ref{2-1 Maya diagrams}) that 
they are given by 
\begin{eqnarray}
&&
\mu(\emptyset,\emptyset,p)=
\left\{
\begin{array}{ll}
(2p-1,2p-2,\ldots,1)   &\ p\geq 1 \\[2mm]
\emptyset              &\ p=0 \\[2mm]
(2|p|,2|p|-1,\ldots,1) &\ p\leq -1~~~.
\end{array}
\right. 
\label{SU(2) ground partition}
\end{eqnarray}
The ground partition with $p\geq 1$ is drawn in Figure 4.
\begin{figure}[tb]
\begin{center}
\includegraphics[scale=0.7]{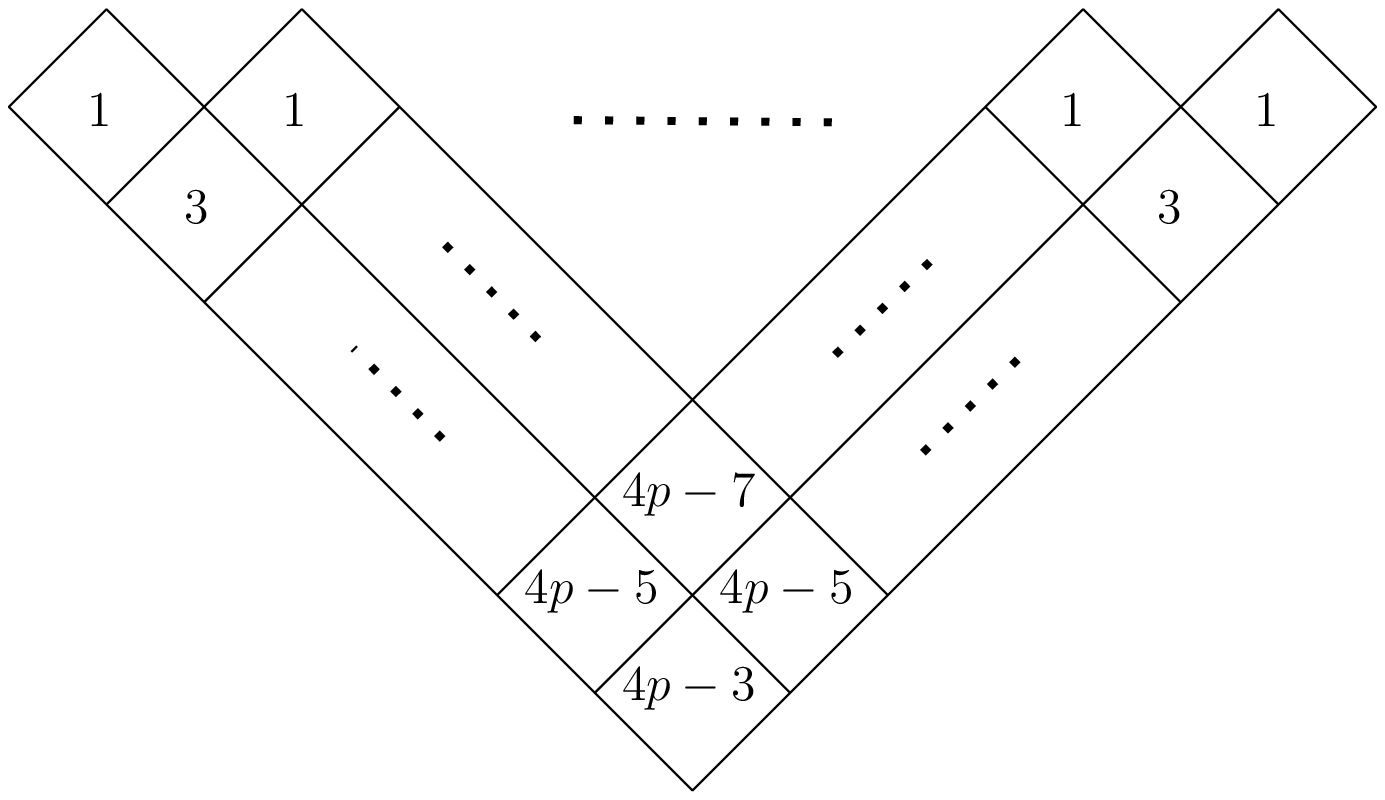}
\mycaption{4}{The $SU(2)$ ground partition for $p \geq 1$.
It is the partition $(2p-1,2p-2,\cdots,\,1)$. 
The numbers in the boxes are the hook length.}
\end{center}\label{N=2 ground partition}
\end{figure}

Taking account of the contributions of 
the ground partitions to (\ref{Z U(1) two-component}), 
we compute the following quantities. 
\begin{eqnarray}
\mathcal{Z}_q^{pert}(p,Q)
& \equiv &
Q^{\frac{1}{2}|\mu(\emptyset,\emptyset;\,p) |}
s_{\mu(\emptyset,\emptyset;\,p)}
(q^{-\rho}).
\label{def Z pert}
\end{eqnarray}
The hook length formula (\ref{hook length formula 2}) 
becomes useful for the computation.  
By plugging the hook length of the boxes (Figure 4) 
into the formula, we evaluate (\ref{def Z pert}) as follows. 
\begin{eqnarray}
Q^{\frac{1}{2}
      \left|\mu (\emptyset,\emptyset;\,p)\right|}
s_{\mu(\emptyset,\emptyset;\,p)}(q^{-\rho})
&=&
Q^{\frac{p(2p-1)}{2}}
\prod_{(i,j)\in \mu(\emptyset,\emptyset;\,p)}
\frac{1}
     {q^{-\frac{h(i,j)}{2}}-q^{\frac{h(i,j)}{2}}}
\nonumber \\
&=&
\prod_{k=1}^{2p-1}
\left\{
\frac{Q^{\frac{1}{2}}}{q^{-\frac{4p-1}{2}}q^k
-q^{\frac{4p-1}{2}}q^{-k}}
\right \}^k \ \ , 
\label{calculation of Zpert in SU(2)}
\end{eqnarray}
for $p\geq 1$. The similar expression is also obtainable 
for $p\leq -1$. We thus get  
\begin{eqnarray}
&&
\mathcal{Z}_q^{pert}(p,Q)=
\left\{
\begin{array}{ll}
\displaystyle{
\prod_{k=1}^{2p-1}
\left\{
\frac{Q^{\frac{1}{2}}}{Q_F^{-\frac{1}{2}}q^k
-Q_F^{\frac{1}{2}}q^{-k}}
\right \}^k }
&\ p\geq 1 \\[0.5cm]
1 
&\ p=0 \\[0.5cm]
\displaystyle{
\prod_{k=1}^{2|p|}
\left\{
\frac{Q^{\frac{1}{2}}}{Q_F^{\frac{1}{2}}q^k
-Q_F^{-\frac{1}{2}}q^{-k}}
\right \}^k }
&\ p\leq -1
\end{array}
\right. \ \ ,
\label{Z pert}
\end{eqnarray} 
where we put $Q_F=q^{4p-1}$.

%%%%%%%%%%%%%%%%%%%%%%%%%%%%%%%%%%%%%%%%%%%%%%
\subsubsection{$SU(2)$ instantons}

We factorize the partition function (\ref{Z U(1) two-component}) 
in the following form.    
\begin{eqnarray}
Z_q^{U(1)}(Q)
&=& 
\sum_{p\in\mathbb{Z}}
\mathcal{Z}_q^{pert}(p,Q)^2
\sum_{\lambda^{(1)},\lambda^{(2)}}
\mathcal{Z}_q^{inst}(\lambda^{(1)},\lambda^{(2)},p,\,Q)^2, 
\label{Z SU(2) factorization}
\end{eqnarray}
where we define $\mathcal{Z}_q^{inst}$ via the relations 
\begin{eqnarray}
Q^{\frac{1}{2} 
   \left|\mu(\lambda^{(1)},\lambda^{(2)};\,p) \right|}
s_{\mu(\lambda^{(1)},\lambda^{(2)};\,p)}(q^{-\rho})
&=&
\mathcal{Z}_q^{pert}(p,Q)\mathcal{Z}_q^{inst}
(\lambda^{(1)},\lambda^{(2)},p,\,Q). 
\label{def Z SU(2) inst}
\end{eqnarray}
Owing to (\ref{def Z pert}) we can rewrite the above as 
\begin{eqnarray}
\mathcal{Z}_q^{inst}(\lambda^{(1)},\lambda^{(2)},p,Q)
&=&
Q^{
\frac{1}{2}
\left(
    |\mu(\lambda^{(1)},\lambda^{(2)};\,p)|
    -|\mu(\emptyset,\emptyset;\,p)|
\right) }
\frac{
  s_{\mu(\lambda^{(1)},\lambda^{(2)};\,p)}(q^{-\rho})} 
{s_{\mu(\emptyset,\emptyset;\,p)}(q^{-\rho})}. 
\label{def Z SU(2) inst schur}
\end{eqnarray}

The ratio of the Schur functions in 
(\ref{def Z SU(2) inst schur}) can be computed 
by using the infinite product formula of 
the (specialized) Schur function \cite{Macdonald} 
\begin{eqnarray}
s_{\mu}(q^{-\rho}) 
=
q^{-\frac{1}{4}\kappa(\mu)}
\prod_{1\leq i<j<\infty}
\left\{
\frac{
q^{-\frac{1}{2}\left(x_i(\mu)-x_j(\mu) \right)}
-q^{\frac{1}{2}\left(x_i(\mu)-x_j(\mu) \right)}}
{q^{-\frac{1}{2}(j-i)}-q^{\frac{1}{2}(j-i)}}
\right \} \  . 
\label{product formula Schur}
\end{eqnarray}
The above formula gives rise to  
the following expression for $\mathcal{Z}_q^{inst}$. 
\begin{eqnarray}
&&
\mathcal{Z}_q^{inst}(\lambda^{(1)},\lambda^{(2)},p,\,Q)
\nonumber \\*[1.5mm]
&&
=
Q^{\frac{1}{2}
\left(
    |\mu(\lambda^{(1)},\lambda^{(2)};\,p)|
    -|\mu(\emptyset,\emptyset;\,p)|
\right) }
q^{-\frac{1}{4}
\left(
\kappa 
   \left(\mu(\lambda^{(1)},\lambda^{(2)};\,p)\right)
-\kappa
   \left(\mu(\emptyset,\emptyset;\,p)\right)
\right)}
\nonumber \\*[1.5mm]
&&
\times 
\prod_{1\leq i<j<\infty}
\left\{
\displaystyle{
\frac{q^{-\frac{1}{2}
        \left(
           x_i(\mu(\lambda^{(1)},\lambda^{(2)};\,p))
           -x_j(\mu(\lambda^{(1)},\lambda^{(2)};\,p)) 
         \right)}
   -q^{\frac{1}{2}
         \left(
            x_i(\mu(\lambda^{(1)},\lambda^{(2)};\,p))
            -x_j(\mu(\lambda^{(1)},\lambda^{(2)};\,p)) 
         \right)}}
{q^{-\frac{1}{2}
         \left(
            x_i(\mu(\emptyset,\emptyset;\,p))
           -x_j(\mu(\emptyset,\emptyset;\,p)) 
          \right)}
    -q^{\frac{1}{2}
         \left(
            x_i(\mu(\emptyset,\emptyset;\,p))
            -x_j(\mu(\emptyset,\emptyset;\,p)) 
         \right)}}
}
\right \}. 
\nonumber \\*[1.5mm]
\label{Z SU(2) inst product}
\end{eqnarray}
This is an expression in terms of 
$\mu(\lambda^{(1)},\lambda^{(2)};\,p)$ and 
$\mu(\emptyset,\emptyset;\,p)$. 
By using (\ref{2-1 Maya diagrams})
we can translate (\ref{Z SU(2) inst product}) 
into an expression presented explicitly in terms 
of $\lambda^{(1)}$ and $\lambda^{(2)}$. 
We first note the simple powers of $Q$ and $q$ in 
(\ref{Z SU(2) inst product}) is read from 
(\ref{2-1 size k}) as  
\begin{eqnarray}
&&
Q^{\frac{1}{2}
\left(
    |\mu(\lambda^{(1)},\lambda^{(2)};\,p)|
    -|\mu(\emptyset,\emptyset;\,p)|
\right) }
q^{-\frac{1}{4}
\left(
\kappa 
   \left(\mu(\lambda^{(1)},\lambda^{(2)};\,p)\right)
-\kappa
   \left(\mu(\emptyset,\emptyset;\,p)\right)
\right)}
\nonumber \\
&&~~~~~~~
=
Q^{|\lambda^{(1)}|+|\lambda^{(2)}|}
q^{-\kappa(\lambda^{(1)})-\kappa(\lambda^{(2)})}
Q_F^{-\frac{|\lambda^{(1)}|-|\lambda^{(2)}|}{2}}. 
\label{Z SU(2) inst simple}
\end{eqnarray}
Due to the correspondence (\ref{2-1 Maya diagrams})   
the elements $x_i(\mu(\lambda^{(1)},\lambda^{(2)};\,p))$ 
and $x_j(\mu(\emptyset,\emptyset;\,p))$ are expressed 
by $x_i(\lambda^{(1)})$, $x_j(\lambda^{(2)})$, 
$x_k(\emptyset)$ and $p$. 
In particular,  
the products in (\ref{Z SU(2) inst product}) 
can be classified into the three patterns according as 
they are made of 
$q^{-\left(x_i(\lambda^{(r)})-x_j(\lambda^{(r)})\right)}
-q^{x_i(\lambda^{(r)})-x_j(\lambda^{(r)})}$ 
($r=1,2$) or 
$Q_F^{-\frac{1}{2}}
q^{-\left(x_i(\lambda^{(1)})-x_j(\lambda^{(2)})\right)}
-Q_F^{\frac{1}{2}}
q^{x_i(\lambda^{(1)})-x_j(\lambda^{(2)})}$. 
We then factor the infinite products into these patterns. 
Putting (\ref{Z SU(2) inst simple}) together we get 
\begin{eqnarray}
&&
\mathcal{Z}_q^{inst}(\lambda^{(1)},\lambda^{(2)},p,Q)
\nonumber \\
&&~~~~
=
\pm 
Q^{|\lambda^{(1)}|+|\lambda^{(2)}|}
q^{-\kappa(\lambda^{(1)})-\kappa(\lambda^{(2)})}
Q_F^{-\frac{|\lambda^{(1)}|-|\lambda^{(2)}|}{2}}
\nonumber \\
&&~~~~~~~~ 
\times 
\prod_{r=1,2}
\left[
\prod_{1\leq i<j<\infty}
\left\{
\frac{
q^{-\big(x_i(\lambda^{(r)})-x_j(\lambda^{(r)})\big)}
        -q^{x_i(\lambda^{(r)})-x_j(\lambda^{(r)})}}
{q^{-(j-i)}-q^{j-i}}
\right\}
\right]
\nonumber \\
&&~~~~~~~~
\times 
\prod_{1\leq i,j<\infty}
\left\{
\frac{
Q_F^{-\frac{1}{2}}
q^{-\big(x_i(\lambda^{(1)})-x_j(\lambda^{(2)})\big)}
-Q_F^{\frac{1}{2}}
q^{x_i(\lambda^{(1)})-x_j(\lambda^{(2)})}}
{Q_F^{-\frac{1}{2}}q^{-(j-i)}
-Q_F^{\frac{1}{2}}q^{j-i}}
\right\}.
\label{Z SU(2) inst for gauge}
\end{eqnarray}

The expression (\ref{Z SU(2) inst for gauge}) 
makes it possible to rewrite $\mathcal{Z}_q^{inst}$ 
in a form convenient to compare with the topological 
string amplitude. 
The first two infinite products have the same form. 
By noting the relation 
$s_{\mu}(q^{-\rho})=(-)^{|\mu|}
q^{-\frac{1}{2}\kappa(\mu)}s_{\mu}(q^{\rho})$, 
the product formula (\ref{product formula Schur}) gives 
\begin{eqnarray}
\prod_{1\leq i<j<\infty}
\left\{
\frac{
q^{-\big(x_i(\lambda^{(r)})-x_j(\lambda^{(r)})\big)}
        -q^{x_i(\lambda^{(r)})-x_j(\lambda^{(r)})}}
{q^{-(j-i)}-q^{j-i}}
\right\}
=
(-)^{|\lambda^{(r)}|}
q^{-\frac{1}{2}\kappa(\lambda^{(r)})}
s_{\lambda^{(r)}}(q^{2\rho}). 
\label{SU(2) infinite product 1}
\end{eqnarray}
As regards the last infinite products in 
(\ref{Z SU(2) inst for gauge}),  
the following description is obtainable 
by the standard computation of 2$d$ CFT. 
\begin{eqnarray}
&&\prod_{1\leq i,j<\infty}
\left\{
\frac{
Q_F^{-\frac{1}{2}}
q^{-\big(x_i(\lambda^{(1)})-x_j(\lambda^{(2)})\big)}
-Q_F^{\frac{1}{2}}
q^{x_i(\lambda^{(1)})-x_j(\lambda^{(2)})}}
{Q_F^{-\frac{1}{2}}q^{-(j-i)}
-Q_F^{\frac{1}{2}}q^{j-i}}
\right\}
\notag\\*[0.5cm]
&&=
q^{\frac{1}{2}
\left(\kappa(\lambda^{(1)})-\kappa(\lambda^{(2)})\right)}
Q_F^{\frac{1}{2}\left(|\lambda^{(1)}|+|\lambda^{(2)}|\right)}
\prod_{k=1}^{\infty}(1-Q_F q^{2k})^{k}
\sum_{\nu}Q_F^{|\nu|}
s_{\nu}(q^{2(\lambda^{(1)}+\rho)})
s_{\nu}(q^{2(\widetilde{\lambda^{(2)}}+\rho)}), 
\nonumber \\*
\label{SU(2) infinite product 2}
\end{eqnarray}
where $\widetilde{\lambda}$ denotes the partition 
conjugate to $\lambda$. Finally, 
putting together (\ref{SU(2) infinite product 1}) 
and (\ref{SU(2) infinite product 2}),  
we can rewrite (\ref{Z SU(2) inst for gauge}) as follows; 
\begin{eqnarray}
&&
\mathcal{Z}_q^{inst}(\lambda^{(1)},\lambda^{(2)},p,Q)
\nonumber \\
&&
~~~~
=
\pm 
Q^{|\lambda^{(1)}|+|\lambda^{(2)}|}
q^{-\kappa(\lambda^{(1)})-2\kappa(\lambda^{(2)})}
Q_F^{|\lambda^{(2)}|} 
s_{\lambda^{(1)}}(q^{2\rho})
s_{\lambda^{(2)}}(q^{2\rho})\notag\\*
&&
~~~~~~~ 
\times 
\prod_{k=1}^{\infty}(1-Q_F q^{2k})^k \ 
\sum_{\nu}Q_F^{|\nu|}
s_{\nu}(q^{2(\lambda^{(1)}+\rho)})
s_{\nu}(q^{2(\widetilde{\lambda^{(2)}}+\rho)}).
\label{Z SU(2) inst for string}
\end{eqnarray}

%%%%%%%%%%%%%%%%%%%%%%%%%%%%%%%%%%%%%%%%%%%%%%%%%
\subsubsection{Interpretation as 
five-dimensional $SU(2)$ Yang-Mills}

We fix $p \in \mathbb{Z}$.  
The relevant field theory parameters are 
$a,\Lambda$ and $R$, 
where $\pm a$ are the VEVs of 
the adjoint scalar in the vector multiplet.
We identify $q, Q_F$ and $Q$ with $a,\Lambda$ and $R$
as follows. 
\begin{eqnarray}
q=e^{-R\hbar},~~~
Q_F=e^{-4Ra},~~~
Q=-(2R\Lambda)^2. 
\label{SU(2) FT parameter}
\end{eqnarray}
Since we have set $Q_F=q^{4p-1}$ the above implies 
\begin{eqnarray}
a=\hbar \widetilde{p},~~~~~ 
\label{SU(2) a and p}
\end{eqnarray}
where 
$\widetilde{p} \equiv p-\frac{1}{4}$.

The identifications (\ref{SU(2) FT parameter})  
convert $\mathcal{Z}_q^{inst}$ 
to the instanton contributions 
in five-dimensional gauge theories. 
This follows by rephrasing 
the expression (\ref{Z SU(2) inst for gauge})
in terms of the field theory parameters. 
In particular, we obtain  
\begin{eqnarray}
\sum_{\lambda^{(1)},\lambda^{(2)}}
\mathcal{Z}_q^{inst}(\lambda^{(1)},\lambda^{(2)},p,Q)^2
&=&
Z^{SU(2)\,inst}_{\,5d\,\mbox{\tiny{SYM}}}(a;\,\Lambda,R,\hbar). 
\label{Z SU(2) SYM inst}
\end{eqnarray}
The RHS is the instanton part of 
the exact partition function \cite{Nekrasov-Okounkov} for 
five-dimensional ${\cal N}=1$ supersymmetric $SU(2)$ Yang-Mills 
with the Chern-Simons term \cite{Chern-Simons}. 
The Chern-Simons corrections come from the following 
factor in (\ref{Z SU(2) inst for gauge}).  
\begin{eqnarray}
q^{-\kappa(\lambda^{(1)})-\kappa(\lambda^{(2)})}
Q_F^{-\frac{|\lambda^{(1)}|-|\lambda^{(2)}|}{2}}, 
\label{SU(2) CS correction}
\end{eqnarray}
which is also understood \cite{Chern-Simons framing} 
as a part of the so-called framing factor 
of the topological string vertices 
\cite{topological vertex}. 
The Chern-Simons term is quantized with 
the coupling constant $c_{cs}$ taking integral values. 
In the present case, it becomes $c_{cs}=2$.

In the next section we will show that 
the square of $\mathcal{Z}_q^{pert}$ is translated 
to the perturbative part of the partition function 
for the supersymmetric Yang-Mills. Taking it for granted for a while, 
together with (\ref{Z SU(2) SYM inst}) we find  
\begin{eqnarray}
\mathcal{Z}_q^{pert}(p,Q)^2
\sum_{\lambda^{(1)},\lambda^{(2)}}
\mathcal{Z}_q^{inst}(\lambda^{(1)}
,\lambda^{(2)},p,Q)^2
&=&
Z^{SU(2)}_{\,5d \, \mbox{\tiny{SYM}}}
(a;\,\Lambda,R,\hbar). 
\label{Z SU(2) SYM}
\end{eqnarray}
The RHS is the exact partition function \cite{Nekrasov-Okounkov} 
for the five-dimensional supersymmetric $SU(2)$ Yang-Mills 
with the Chern-Simons term.  
We notice that, 
as confirmed \cite{Nekrasov-Okounkov} 
for four-dimensional $\mathcal{N}=2$ 
supersymmetric gauge theories,  
the gauge theory partition function is realized as 
the $\hbar \rightarrow 0$ limit of 
the above partition function.

It is clear from (\ref{Z SU(2) SYM}) that the field theory 
limit is the thermodynamic limit with $a, \Lambda$ and $R$ fixed. 
The identification (\ref{SU(2) a and p}) 
makes $p=o(\hbar^{-1})$. 
The corresponding ground partition in 
(\ref{Z SU(2) SYM}) becomes very large. 
The partitions $\lambda^{(r)}$ represent 
$U(1)$ gauge instantons with $c_2=|\lambda^{(r)}|$ 
on a non-commutative $\mathbb{R}^4$. 
In the model of random partitions,  
it follows from (\ref{2-1 Maya diagrams}) that 
these gauge instantons become excitations 
from the ground partition.  
They provide the deformation of the ground partition 
at the thermodynamic limit (Figure 5).  
These imply that the ground partitions 
describe the perturbative regime of the Coloumb branch 
while their deformation becomes non-perturbative in the gauge theory.  
\begin{figure}[h]
\begin{center}
\includegraphics[scale=0.5]{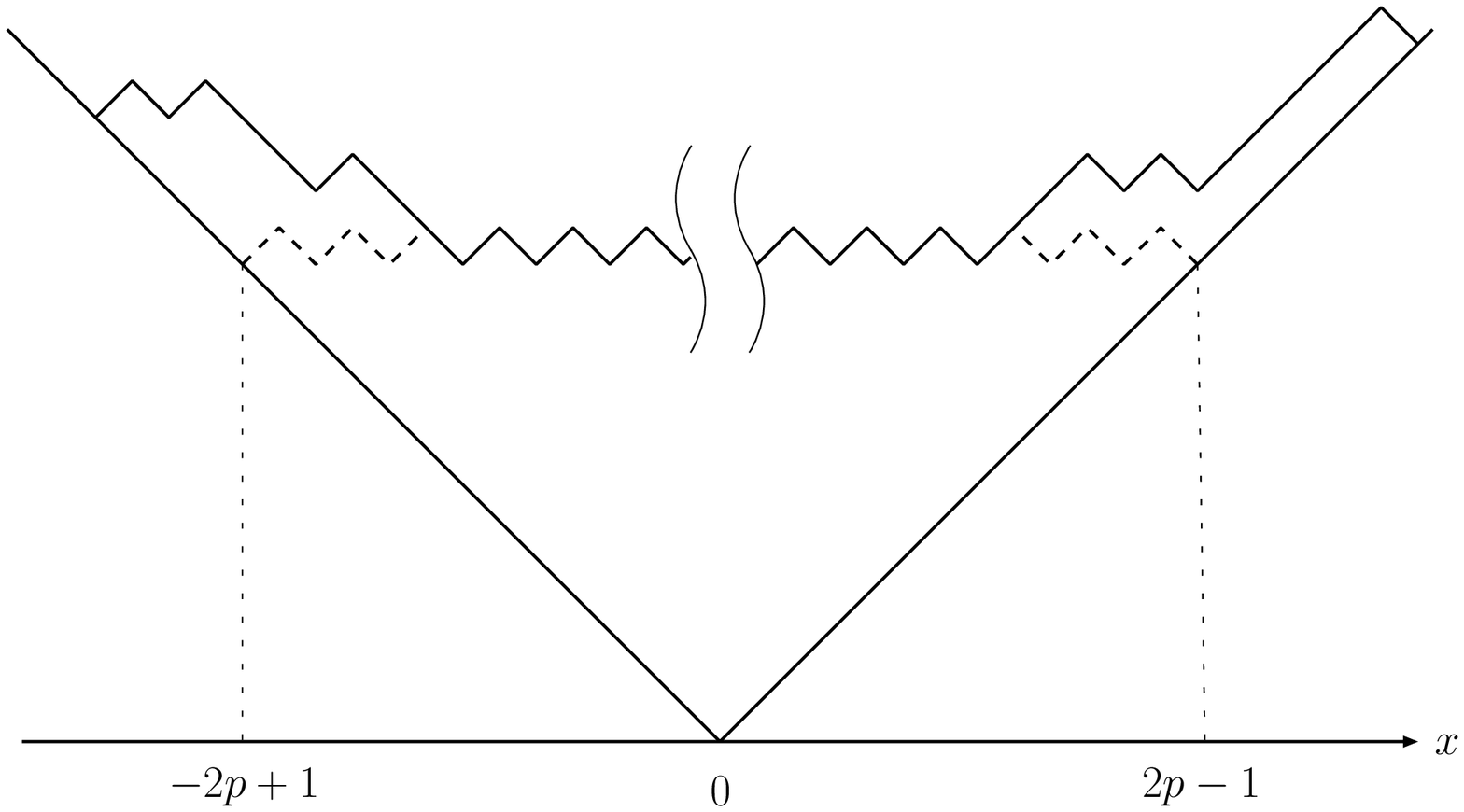}
\mycaption{5}{Deformation of the $SU(2)$ ground partition 
for $p \gg 1$ by two partitions $(3,1,1)$ and 
$(3,2)$.}
\end{center}\label{pileup}
\end{figure}
%
%

%%%%%%%%%%%%%%%%%%%%%%%%%%%%%%%%%%%%%%%%%%%%%%%%%
\subsection{Interpretation as topological string amplitude}

The relevant non-compact Calabi-Yau threefold 
is the local $\mathbb{F}_2$. 
The geometric engineering \cite{Geometric engineering} 
dictates that four-dimensional $\mathcal{N}=2$ supersymmetric 
$SU(2)$ Yang-Mills is realized by the canonical bundle over 
the Hirzebruch surface $\mathbb{F}_m$. 
It is an ALE space with $A_1$ singularity fibred  
over $\mathbb{P}^1$. 
Type of the fibrations of the ALE space 
is labelled by the integer $m$, which is  
called the framing and taking values $0,1,2$. 
We choose the framing to be $m=2$. 
The geometrical data are the K\"ahler volumes 
of the compact two-cycles. 
We denote the K\"ahler parameters 
of the base $\mathbb{P}^1$ 
and the fibre $\mathbb{P}^1$ 
(the blow-up cycle of $A_1$ singularity) 
respectively by $t_B$ and $t_F$.

Topological string amplitudes on the geometries 
dictated by the geometric engineering are computed 
in \cite{Iqbal,Eguchi} by using the topological vertices.  
We summarize these amplitudes in Appendix. 
Since they are generating functions of 
the world-sheet instantons, string coupling constant 
$g_{st}$ and the K\"ahler parameters appear 
in the amplitudes as certain combinations 
of $q=e^{-g_{st}}$,$Q_B=e^{-t_B}$ and $Q_F=e^{-t_F}$.

Let us denote the topological string amplitude 
on the local $\mathbb{F}_2$ by  
$Z^{SU(2)}_{string}(q,Q_F,Q_B)$. 
It is given in (\ref{SU(2) string amplitude}). 
By comparing (\ref{Z SU(2) inst for string}) with 
(\ref{SU(2) string amplitude}), we see that 
the gauge instanton contribution 
(\ref{Z SU(2) SYM inst}) can be converted 
to the string amplitude as follows.
\begin{eqnarray}
\sum_{\lambda^{(1)},\lambda^{(2)}}
\mathcal{Z}_q^{inst}(\lambda^{(1)}
,\lambda^{(2)},p,Q)^2
&=&
\prod_{k=1}^{\infty}
(1-Q_F q^{2k})^{2k}\,\,
Z_{string}^{SU(2)}(q^2,Q_F,Q^2), 
\label{SU(2) ins vs string}
\end{eqnarray}
where we put $Q_F=q^{4\widetilde{p}}$.  
This is in accord with the result 
\cite{Iqbal,Eguchi} that 
the topological string amplitude 
on the local $\mathbb{F}_2$ leads to 
the instanton part of the partition function 
of five-dimensional supersymmetric $SU(2)$ Yang-Mills.

Due to the identification $\hbar \sim g_{st}$,  
the thermodynamic limit is in the perturbative regime 
of string theory. The previous consideration about  
the role of the ground partitions in gauge theory 
could be translated in string theory. 
The ground partitions at the thermodynamic limit 
must be interpreted as classical objects in string theory. 
The relation $t_F=4Ra$, 
which follows from (\ref{SU(2) ins vs string}), suggests 
that the ground partitions describe 
the resolution of $A_1$ singularity 
in the local $\mathbb{F}_2$.

%%%%%%%%%%%%%%%%%%%%%%%%%%%%%%%%%%%%%%%%%%%%%%%%%%%%%%%%%%
%%%%%%%%%%%%%%%%%%%%%%%%%%%%%%%%%%%%%%%%%%%%%%%%%%%%%%%%%%
%%%%%%%%%%%%%%%%%%%%%%%%%%%%%%%%%%%%%%%%%%%%%%%%%%%%%%%%%%
\section{Perturbative gauge theory from ground partitions}
\label{section4}

The aim of this section is to show that the square 
of $\mathcal{Z}_{q}^{pert}(p,Q)$ in 
(\ref{Z SU(2) SYM}) describes the perturbative part 
of the partition function for 
the five-dimensional supersymmetric Yang-Mills.

Let $\mathcal{G}_{n}$ be the partition $(n,n-1,\cdots,1)$, 
where $n$ is a positive integer.  
The ground partition $\mu(\emptyset,\emptyset;\,p)$ 
for $p \in \mathbb{Z}_{\geq 1}$ is $\mathcal{G}_{2p-1}$  
\footnote{
The ground partition 
for $p \in \mathbb{Z}_{\leq -1}$
is $\mathcal{G}_{|2p|}$. 
The discussion presented below 
goes as well in these cases.}.  
Let us consider the logarithm of 
$\mathcal{Z}_{q}^{pert}(p,Q)^{2}$. 
It can be written as follows. 
\begin{eqnarray}
-\ln 
\mathcal{Z}_{q}^{pert}(p,Q)^{2} 
&=& 
\ln 
\prod_{(i,j) \in  \mathcal{G}_{2p-1}}
\frac{\left( q^{-\frac{h(i,j)}{2}}
        -q^{\frac{h(i,j)}{2}} \right)^{2}}
     {Q}
\notag\\
&=& 
\ln 
\prod_{k=1}^{2p-1}
\frac{\left(  e^{R\hbar(2\tilde{p}-k)}
         -e^{-R\hbar(2\tilde{p}-k)} \right)^{2k}}
        {(-)^k(2R\Lambda)^{2k}}, 
\label{ln pert}
\end{eqnarray} 
where $q$ and $Q$ are translated to $\Lambda$ and $R$ 
by (\ref{SU(2) FT parameter}). 
We also note $\widetilde{p}=p-\frac{1}{4}$. 
We further factor the logarithm  as  
\begin{eqnarray} 
&&
-\ln \mathcal{Z}_{q}^{pert}(p,Q)^{2} 
\nonumber \\
&&~~~~
=\sum_{k=1}^{2p-1}
k \ln 
\left( \frac{e^{R\hbar(2\tilde{p}-k)}
              -e^{-R\hbar(2\tilde{p}-k)}}
            {2R\Lambda} \right)
+\sum_{k=1}^{2p-1}
k \ln 
\left(  \frac{e^{-R\hbar(2\tilde{p}-k)}
               -e^{R\hbar(2\tilde{p}-k)}}
             {2R\Lambda} \right)   
\nonumber \\
&&~~~~ 
=g(2p\,|\hbar,2R,\Lambda)+g(2p\,|-\hbar,2R,\Lambda), 
\label{ln Z by g}
\end{eqnarray}
where we introduce  
\begin{eqnarray}
g(2p\,|\hbar,2R,\Lambda)
&=&
\sum_{k=1}^{2p-1}
k \ln 
\left(
   \frac{ 
     \sinh{R\hbar(2\tilde{p}-k)}}{R\Lambda}
\right).
\label{g}
\end{eqnarray}

The function (\ref{g}) becomes identical to 
the function $\gamma_{\hbar}(x;\beta;\Lambda)$ 
given in \cite{Nekrasov-Okounkov}. To confirm this, 
it should be tested first of all 
by the q-difference equation 
\cite{Nekrasov-Okounkov}.

\begin{proposition}
\label{q-difference equation}
The function (\ref{g}) satisfies 
the q-difference equation
\begin{eqnarray}
g(2p+1|\hbar,2R,\Lambda)+g(2p-1|\hbar,2R,\Lambda)
-2g(2p|\hbar,2R,\Lambda)
&=&
\ln\left(
\frac{\sinh R(2\hbar\tilde{p})}{R\Lambda}\right).
\label{q-diff}
\end{eqnarray}
\end{proposition}

Let us verify (\ref{q-diff}) by using a recurrence relation  
among the Young diagrams $\mathcal{G}_n$. 
The Young diagrams $\mathcal{G}_{2p}$ and $\mathcal{G}_{2p-2}$ 
are piled in Figure 6-(a) so that any two boxes 
at a common position have the same hook length.  
While in Figure 6-(b),  
two $\mathcal{G}_{2p-1}$ are slightly shifted 
to cover the piled diagrams of (a).  
Each box of the diagrams has the same hook length 
as that of $\mathcal{G}_{2p}$ at the same position. 
By noting (\ref{ln pert}), we exponentiate 
the LHS of (\ref{q-diff}) to 
\begin{eqnarray}
&&
\prod_{(i,j)\in \mathcal{G}_{2p}}
(q^{-\frac{h(i,j)}{2}}-q^{\frac{h(i,j)}{2}})
\prod_{(i,j)\in \mathcal{G}_{2p-1}}
(q^{-\frac{h(i,j)}{2}}-q^{\frac{h(i,j)}{2}})^{-1}
\notag\\
&&
\qquad\qquad\qquad\qquad
\times\prod_{(i,j)\in \mathcal{G}_{2p-1}}
(q^{-\frac{h(i,j)}{2}}-q^{\frac{h(i,j)}{2}})^{-1}
\prod_{(i,j)\in \mathcal{G}_{2p-2}}
(q^{-\frac{h(i,j)}{2}}-q^{\frac{h(i,j)}{2}}). 
\label{q-difference equation diagrams}
\end{eqnarray}
It follows from the above explanation of the figures 
that nearly all the products which come 
from $\mathcal{G}_{2p}$ and $\mathcal{G}_{2p-2}$ 
in (\ref{q-difference equation diagrams}) 
cancel with the products coming 
from two $\mathcal{G}_{2p-1}$, and 
that the net becomes only 
the contribution from the shaded box 
in Figure 6-(c). 
Hence, 
(\ref{q-difference equation diagrams}) 
is equal to 
$q^{-\frac{4p-1}{2}}-q^{\frac{4p-1}{2}}$. 
This turns to give the RHS of (\ref{q-diff}). 
\begin{figure}[h]
\begin{center}
\includegraphics[scale=0.5]{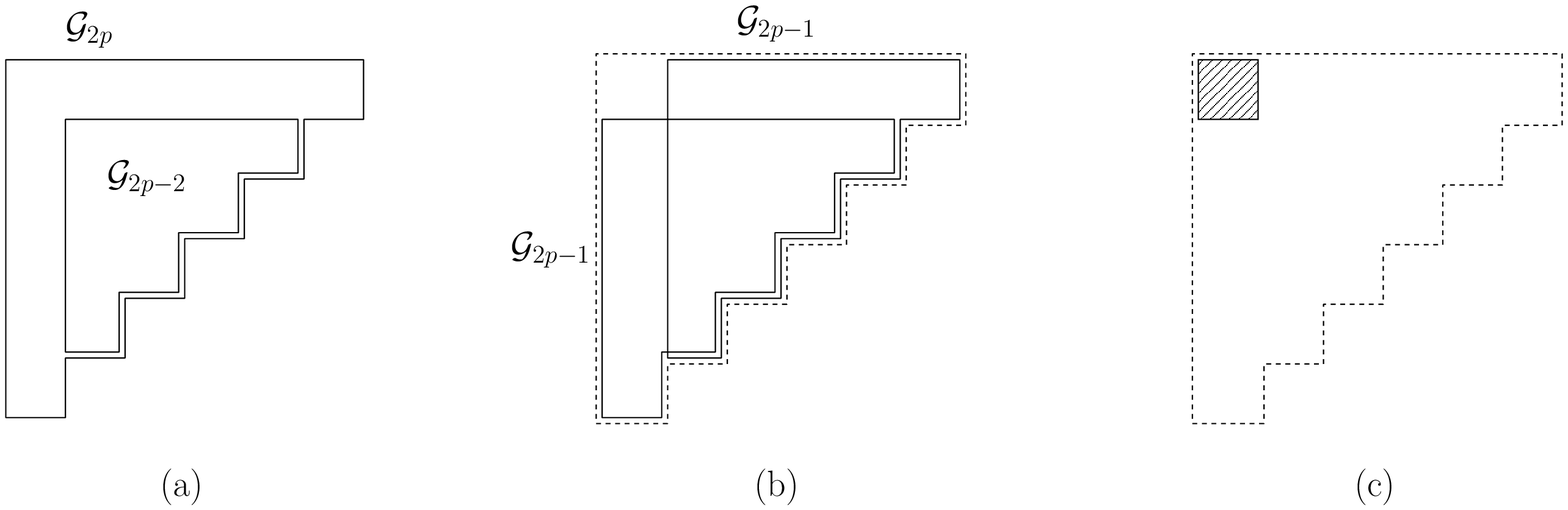}
\mycaption{6}{The Young diagrams for 
the $q$-difference equation.} 
\end{center}\label{Q-difference}
\end{figure}

\begin{proposition}
\label{gamma N-O}
The function $g(2p|\hbar,2R,\Lambda)$ coincides 
with the function 
$\gamma_{\hbar}(2\hbar\tilde{p}|2R;\Lambda)$ 
in \cite{Nekrasov-Okounkov},  
up to a linear function in $\hbar\tilde{p}$.  
\end{proposition}

To prove the above we rewrite (\ref{g}) as 
\begin{eqnarray}
&&
g(2p|\hbar,2R,\Lambda)
\nonumber \\
&&~~~~
=
-\sum_{k=1}^{2p-1}
k \ln (2R\Lambda)
+R\hbar\sum_{k=1}^{2p-1}k(2\tilde{p}-k)
+\sum_{k=1}^{2p-1}
k \ln \left(1-e^{-2R\hbar(2\tilde{p}-k)}\right). 
\label{hook}
\end{eqnarray}
The first two terms become 
\begin{eqnarray}
-\sum_{k=1}^{2p-1}
k \ln (2R\Lambda) 
&=&
-\frac{(2\hbar\widetilde{p})^2}{2\hbar^2} 
\ln (2R\Lambda)
+
\frac{1}{8}
\ln (2R \Lambda), 
\nonumber \\*
R\hbar 
\sum_{k=1}^{2p-1}
k(2\widetilde{p}-k)
&=& 
\frac{R
(2\hbar \widetilde{p})^3}
{6\hbar^2}
-
\frac{R
(2\hbar \widetilde{p})}
{24}.
\label{polynomial of g}
\end{eqnarray}
As regards the last term of (\ref{hook})   
we rewrite the finite sum as 
a subtraction between two infinite sums 
and then expand the logarithms 
\begin{eqnarray}
&&
\sum_{k=1}^{2p-1}
k\ln\left(1-e^{-2R\hbar(2\tilde{p}-k)}\right) 
= 
\left\{
\sum_{k=1}^{+\infty}
-\sum_{k=2p}^{+\infty}
\right\}
k\ln\left(1-e^{-2R\hbar(2\tilde{p}-k)}\right) 
\nonumber \\
&&~~~~~~~~~
=
-\sum_{k=1}^{\infty}
k\sum_{m=1}^{\infty}
\frac{1}{m}e^{-2Rm\hbar(2\tilde{p}-k)}
+\sum_{k=2p}^{\infty}k\sum_{m=1}^{\infty}
\frac{1}{m}e^{-2Rm\hbar(2\tilde{p}-k)}.
\label{for ln g}
\end{eqnarray}
The summations over $k$ in the above 
give rise to 
\begin{eqnarray}
&& 
\sum_{k=1}^{2p-1}
k\ln(1-e^{-2Rm\hbar(2\tilde{p}-k)})
\nonumber \\
&&~~~~~ 
= \sum_{m=1}^{\infty}\frac{1}{m}
\frac{e^{-Rm(2\hbar\tilde{p})}}{(1-e^{2Rm\hbar})
(1-e^{-2Rm\hbar})} 
\nonumber \\
&&~~~~~~~~~~
+ 
(\widetilde{p}+\frac{1}{4})
\sum_{m=1}^{\infty}
\frac{1}{m}\frac{e^{Rm\hbar}}{1-e^{2Rm\hbar}} 
-\sum_{m=1}^{\infty}
\frac{1}{m}
\frac{e^{Rm\hbar}}{(1-e^{2Rm\hbar})(1-e^{-2Rm\hbar})},  
\label{ln of g}
\end{eqnarray}
where the first term provides 
an analogue of the McMahon function,  
and the last two terms are a linear function 
in $\hbar\tilde{p}$.

Collecting (\ref{polynomial of g}) and (\ref{ln of g}),  
we obtain the following expression for 
$g(2p|\hbar,2R,\Lambda)$. 
\begin{eqnarray}
&&
g(2p|\hbar,2R,\Lambda) 
\nonumber \\
&&~~~~~~~
= 
\alpha(\hbar,2R,\Lambda)
+\beta(\hbar,2R,\Lambda)(\hbar\tilde{p})
\nonumber \\
&& ~~~~~~~~~~~
-\frac{(2\hbar\tilde{p})^{2}}{2\hbar^{2}}
\ln (2R\Lambda)
+\frac{R(2\hbar\tilde{p})^{3}}{6\hbar^{2}}
+\sum_{m=1}^{\infty}\frac{1}{m}
\frac{e^{-2Rm(2\hbar\tilde{p})}}
{(1-e^{2Rm\hbar})(1-e^{-2Rm\hbar})}
\nonumber \\
&&~~~~~~~
=
\alpha(\hbar,2R,\Lambda)
+\beta(\hbar,2R,\Lambda)(\hbar\tilde{p})
+\gamma_{\hbar}(2\hbar\tilde{p}|2R,\Lambda).
\label{vs.gamma1}
\end{eqnarray}
Thus up to a linear function in $\hbar\tilde{p}$, 
the function (\ref{g}) is identified with 
the perturbative term given 
in \cite{Nekrasov-Okounkov}. 
We lastly remark that 
$g(p|\hbar,2R,\Lambda)$ has 
the smooth $R \to 0$ limit, 
as is obvious from (\ref{g}).

By the same argument as above 
we can also express $g(2p|-\hbar,2R,\Lambda)$ 
in a form analogous to (\ref{vs.gamma1}).  
It becomes as follows. 
\begin{eqnarray}
&&
g(2p|-\hbar,2R,\Lambda) 
\nonumber \\
&&~~~~~~~
= \alpha(-\hbar,2R,\Lambda)
+\beta(-\hbar,2R,\Lambda)(-\hbar\tilde{p})
\nonumber \\
&&~~~~~~~~~~~~
-\frac{(2\hbar\tilde{p})^{2}}{2\hbar^{2}}
\ln(2R\Lambda)
-\frac{R(2\hbar\tilde{p})^{3}}{6\hbar^{2}}
+\sum_{m=1}^{\infty}
\frac{1}{m}
\frac{e^{2Rm(2\hbar\tilde{p})}}{(1-e^{2Rm\hbar})
(1-e^{-2Rm\hbar})}
\nonumber \\
&&~~~~~~~
=\alpha(-\hbar,2R,\Lambda)
+\beta(-\hbar,2R,\Lambda)(-\hbar\tilde{p})
+\gamma_{h}(-2\hbar\tilde{p}|R,\Lambda).
\label{vs.gamma2}
\end{eqnarray}
Combining (\ref{vs.gamma1}) and  (\ref{vs.gamma2}), 
we arrive at
\begin{eqnarray}
\mathcal{Z}_{q}^{pert}(p,Q)^{2} 
&=& 
\exp 
\Bigl\{
-g(2p|\hbar,2R,\Lambda)
-g(2p|-\hbar,2R,\Lambda)
\Bigr\}
\nonumber \\
&=& 
\exp 
\Bigl \{
-\Delta^{(0)}(\hbar,2R,\Lambda)
-\Delta^{(1)}(\hbar,2R,\Lambda)
(\hbar \widetilde{p})
\Bigr \}
\nonumber \\
&&
\times 
\exp 
\left \{
  -\Bigl (
       \gamma_{\hbar}(2\hbar\tilde{p}|2R,\Lambda)
       +\gamma_{\hbar}(-2\hbar\tilde{p}|2R,\Lambda)
   \Bigr )
\right \}, 
\label{vs.pert}
\end{eqnarray}
where 
$\Delta^{(0)}$ and $\Delta^{(1)}$ 
are the collections of $\alpha$ and $\beta$ 
in (\ref{vs.gamma1}) and (\ref{vs.gamma2}). 
In terms of the VEV of the adjoint scalar 
(\ref{SU(2) a and p}) the above becomes 
\begin{eqnarray}
\mathcal{Z}_{q}^{pert}(p,Q)^{2} 
&=& 
\exp 
\Bigl \{
-\Delta^{(0)}(\hbar,2R,\Lambda)
-\Delta^{(1)}(\hbar,2R,\Lambda)a
\Bigr \}
\nonumber \\
&&
\times 
\exp 
\left \{
  -\Bigl (
       \gamma_{\hbar}(2a|2R,\Lambda)
       +\gamma_{\hbar}(-2a|2R,\Lambda)
   \Bigr )
\right \}.
\label{vs.pert 2}
\end{eqnarray}
The expression (\ref{vs.pert 2}) shows that 
the square of $\mathcal{Z}_{q}^{pert}$ 
can be identified with the perturbative part of 
the partition function for the five-dimensional 
supersymmetric Yang-Mills.

%%%%%%%%%%%%%%%%%%%%%%%%%%%%%%%%%%%%%%%%%%%%%%%%%%%%%%%%%%%%
%%%%%%%%%%%%%%%%%%%%%%%%%%%%%%%%%%%%%%%%%%%%%%%%%%%%%%%%%%%%
\section{$SU(N)$ Yang-Mills and random plane partitions}
\label{section5}

The previous discussions on the $SU(2)$ gauge theory 
could be generalized to the higher rank gauge groups. 
Owing to the identification (\ref{N-1}) of 
$N$-component fermions 
any charged partition $(\mu,n)$ can be expressed 
by a set of $N$ charged partitions $(\lambda^{(r)},p_r)$.   
We consider in the neutral sector, 
that is, $n=\sum_{r=1}^Np_r=0$. 
$(\lambda^{(r)},p_r)$  determine a partition 
$\mu \left(\{\lambda^{(r)}\};\,\{p_r\}\right)$
in this sector. 
The corresponding Maya diagram 
$S_{\mu \left(\{\lambda^{(r)}\};\,\{p_r\} \right)}$ 
can be read from (\ref{N-1 Maya diagrams}) as 
\begin{eqnarray}
S_{\mu \left(\{\lambda^{(r)}\};\,\{p_r\} \right)}
&=&
\bigcup_{r=1}^N~
\Bigl \{
N(x_{i_r}(\lambda^{(r)})+\widetilde{p}_r)~ 
;i_r \geq 1 
\Bigr \}.
\label{SU(N) Maya diagrams}
\end{eqnarray}

In this setting 
it is convenient to colour-code a partition. 
We $N$-colour the upper edges of the Young diagram 
by attaching $r$ to the edge 
when the projection of the middle point 
belongs to $N\mathbb{Z}+r-\frac{1}{2}$. 
Thereby rows and columns of the Young diagram 
are $N$-coloured. 
See Figure 7 for the case of $N=3$. 
The box $(i,j)$ in the Young diagram 
is bicoloured by $(r_i,s_j)$, where $r_i$ and $s_j$ are 
respectively the colors of the $i$-th row and the $j$-th 
column. The box bicoloured by $(r,s)$ is simply called  
$(r,s)$-box.

We regard the random plane partitions (\ref{Z U(1)}) 
as the $q$-deformed random partitions 
via (\ref{Z U(1) schur}) 
and rewrite the partition function as 
\begin{eqnarray}
Z_q^{U(1)}(Q)
&=&
\sum_{ \{ p_r \} } 
\sum_{ \{ \lambda^{(r)} \}} 
Q^{\left| \mu \left(\{\lambda^{(r)}\};\,\{p_r\} \right) 
\right|}
s_{\mu \left( \{\lambda^{(r)}\};\,\{p_r\} \right)}
(q^{-\rho})^2.
\label{Z U(1) N-component}
\end{eqnarray}
Similarly to the case of $SU(2)$, we will further 
factor the above into the following form. 
\begin{eqnarray}
Z_q^{U(1)}(Q)
&=&
\sum_{\{p_r\}}
Z_q^{SU(N)\,pert}(\{p_r\},Q)\,
\sum_{\{\lambda^{(r)}\}}
Z_q^{SU(N)\,inst}
(\{\lambda^{(r)}\},\{p_r\},Q).  
\label{Z SU(N) factorization pre}
\end{eqnarray}

%%%%%%%%%%%%%%%%%%%%%%%%%%%%%%%%%%%%%%%%%%%%%%
\subsubsection*{$SU(N)$ ground partitions}

We consider the cases of the empty partitions    
\begin{eqnarray}
\lambda^{(r)}= \emptyset~~~~
(1 \leq {\forall}r \leq N). 
\end{eqnarray}
For each $\{p_r\}$, 
the corresponding partition 
$\mu \left( \{\emptyset \};\{p_r\} \right)$ 
is named the $SU(N)$ ground partition.  
The case of $N=3$ is illustrated in Figure 7. 
\begin{figure}[htb]
\begin{center}
\includegraphics[scale=0.6]{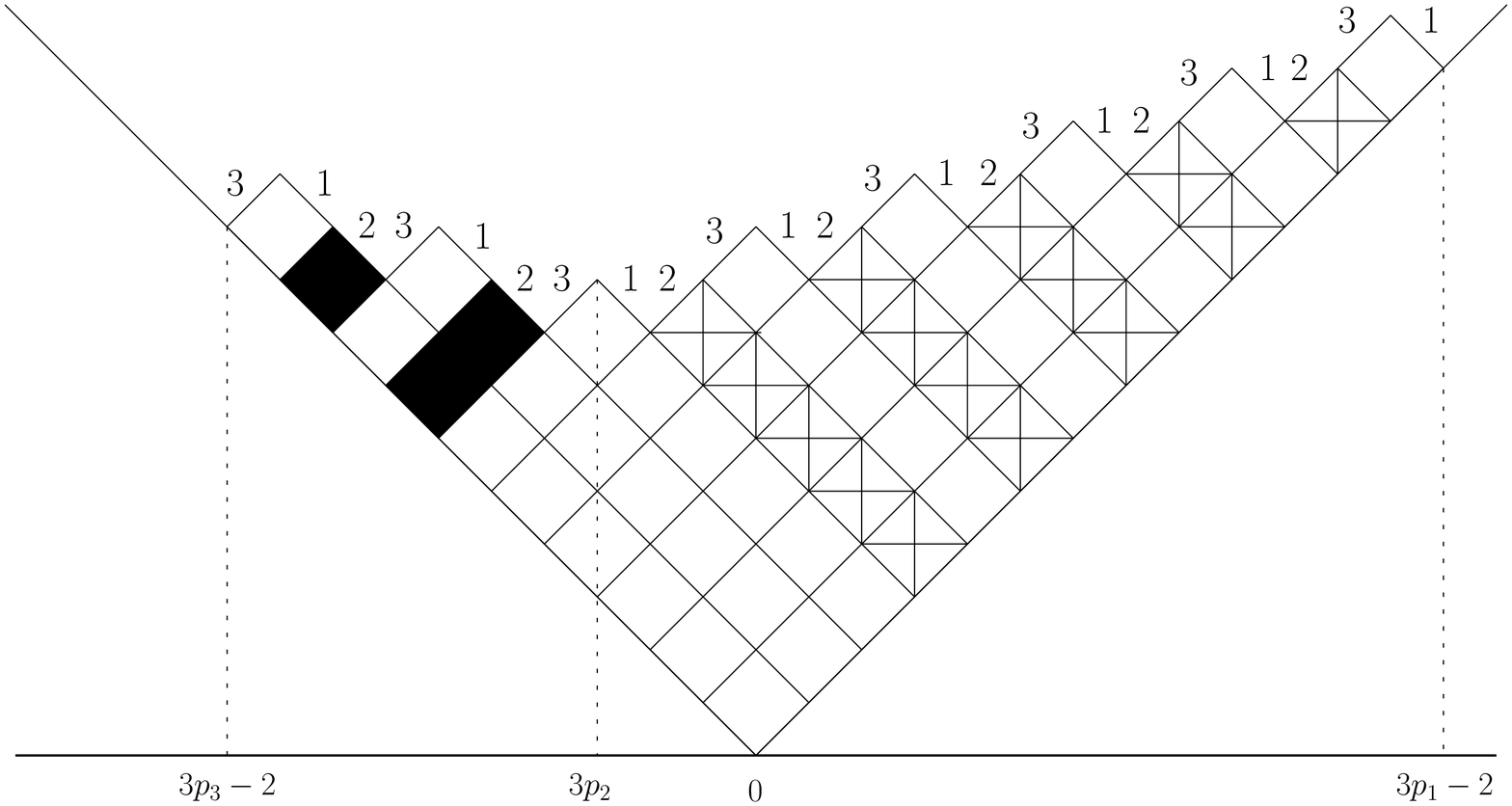}
\mycaption{7}{The $SU(3)$ ground partition for 
$(p_1,p_2,p_3)=(5,-1,-4)$.
The rows and the columns are coloured 
by $1,2$ and $3$. The Young diagram 
is bicoloured by $(1,2), (1,3)$ and $(2,3)$. 
The corresponding boxes are denoted respectively 
by $\boxtimes$, $\square$ and $\blacksquare$.}
\end{center}\label{N=3 ground partition}
\end{figure}
Their contributions to the partition function 
lead us to introduce 
$\mathcal{Z}_q^{SU(N)\,pert}(\{p_r\},Q)$ by 
\begin{eqnarray}
q^{-\frac{N^2}{12}\sum_{s=1}^{N}\widetilde{p_s}^3}
\mathcal{Z}_q^{SU(N)\,pert}(\{p_r\},Q)
&\equiv& 
Q^{\frac{1}{2}
  |\mu\left(\{\emptyset \};\,\{p_r\}\right)|}
s_{\mu\left(\{\emptyset \};\,\{p_r\}\right)}
(q^{-\rho}).  
\label{def Z SU(N) pert}
\end{eqnarray}
In other words, we put   
\begin{eqnarray}
\mathcal{Z}_q^{SU(N)\,pert}(\{p_r\},Q)
&=& 
\prod_{(i,j)
  \in \mu \left(\{\emptyset \};\,\{p_r\}\right)}
\frac{  Q^{\frac{1}{2}} }
     {  q^{-\frac{h(i,j)}{2}}
       -q^{ \frac{h(i,j)}{2}} }. 
\label{def Z SU(N) pert hook}
\end{eqnarray}
This follows from 
(\ref{def Z SU(N) pert}) by applying 
the formula (\ref{hook length formula 2}) and noting 
\footnote{$\kappa(\mu)$ measures asymmetry of the Young diagram.}
$\kappa(\mu(\{\emptyset\};\{p_r\}))
=\frac{N^2}{3}\sum_{s=1}^N \widetilde{p}_s^3$.

Let us compute $\mathcal{Z}_q^{SU(N)\,pert}$ 
given in the above. 
It is instructive to start with the case of $N=3$. 
To simplify the discussion, 
we order the three $U(1)$ charges $p_{1,2,3}$ 
as $p_1>p_2>p_3$. 
They satisfy $p_1+p_2+p_3=0$ by the neutral condition.  
The ground partition for $(p_1,p_2,p_3)$ 
is bicoloured by $(1,2),(1,3)$ and $(2,3)$. 
We factor the products in (\ref{def Z SU(N) pert hook}) 
according to the color-coding of the partition as follows. 
\begin{eqnarray}
&&
\prod_{(i,j)\in\mu(\{\emptyset\};\{p_1,p_2,p_3\})}
\left(q^{-\frac{h(i,j)}{2}}
-q^{\frac{h(i,j)}{2}}\right)^{-1}
\notag\\*
&&\ \ \ \ 
=\prod_{\boxtimes\in\mu(\{\emptyset\};\{p_1,p_2,p_3\})}
\left(q^{-\frac{h(\boxtimes)}{2}}
-q^{\frac{h(\boxtimes)}{2}}\right)^{-1}
\prod_{\square\in\mu(\{\emptyset\};\{p_1,p_2,p_3\})}
\left(q^{-\frac{h(\square)}{2}}
-q^{\frac{h(\square)}{2}}\right)^{-1}
\notag\\*
&& \ \ \ \ \ \ \ 
\times 
\prod_{\blacksquare\in\mu(\{\emptyset\};\{p_1,p_2,p_3\})}
\left(q^{-\frac{h(\blacksquare)}{2}}
-q^{\frac{h(\blacksquare)}{2}}\right)^{-1}, 
\label{factor N=3 grand hook}
\end{eqnarray}
where we depict $(1,2)$-box as $\boxtimes$, 
$(1,3)$-box as $\square$ and $(2,3)$-box 
as $\blacksquare$. 
Each term of (\ref{factor N=3 grand hook}) 
can be computed so that it acquires a form  
analogous to (\ref{calculation of Zpert in SU(2)}).
For instance, the first term becomes 
\begin{eqnarray}
\prod_{\boxtimes\in\mu(\{\emptyset\};\{p_1,p_2,p_3\})}
\left(q^{-\frac{h(\boxtimes)}{2}}
-q^{\frac{h(\boxtimes)}{2}}\right)^{-1}
&=& 
\prod_{k=1}^{p_1-p_2-1}
\left(q^{-\frac{3}{2}
(\widetilde{p}_1-\widetilde{p}_2-k)}
-q^{\frac{3}{2}
(\widetilde{p}_1-\widetilde{p}_2-k)}\right)^{-k}. 
\label{component N=3 grand hook}
\end{eqnarray}
These give the following expression for 
$\mathcal{Z}_q^{SU(3)\,pert}$;  
\begin{eqnarray}
\mathcal{Z}_q^{SU(3)\,pert}(p_1,p_2,p_3,Q)=
\prod_{1\leq r<s\leq 3}\prod_{k=1}^{p_r-p_s-1}
\left\{ 
\frac{Q^{\frac{1}{2}}}
{q^{-\frac{3}{2}
      (\widetilde{p}_r-\widetilde{p}_s-k)}
 -q^{\frac{3}{2}
       (\widetilde{p}_r-\widetilde{p}_s-k)}}
\right \}^k. 
\label{Z SU(3) pert}
\end{eqnarray}

For the cases of general values of $N$, 
we also order the $U(1)$ charges as 
$p_1 > p_2 > \cdots > p_N$. 
They satisfy $\sum_{r=1}^Np_r=0$. 
The $SU(N)$ ground partition for $\{p_r \}$ 
is now bicoloured by $(r,s)$ with 
$1 \leq r < s \leq N$. See Figure 8 
for the illustration. 
\begin{figure}[t]
\begin{center}
\includegraphics[scale=0.7]{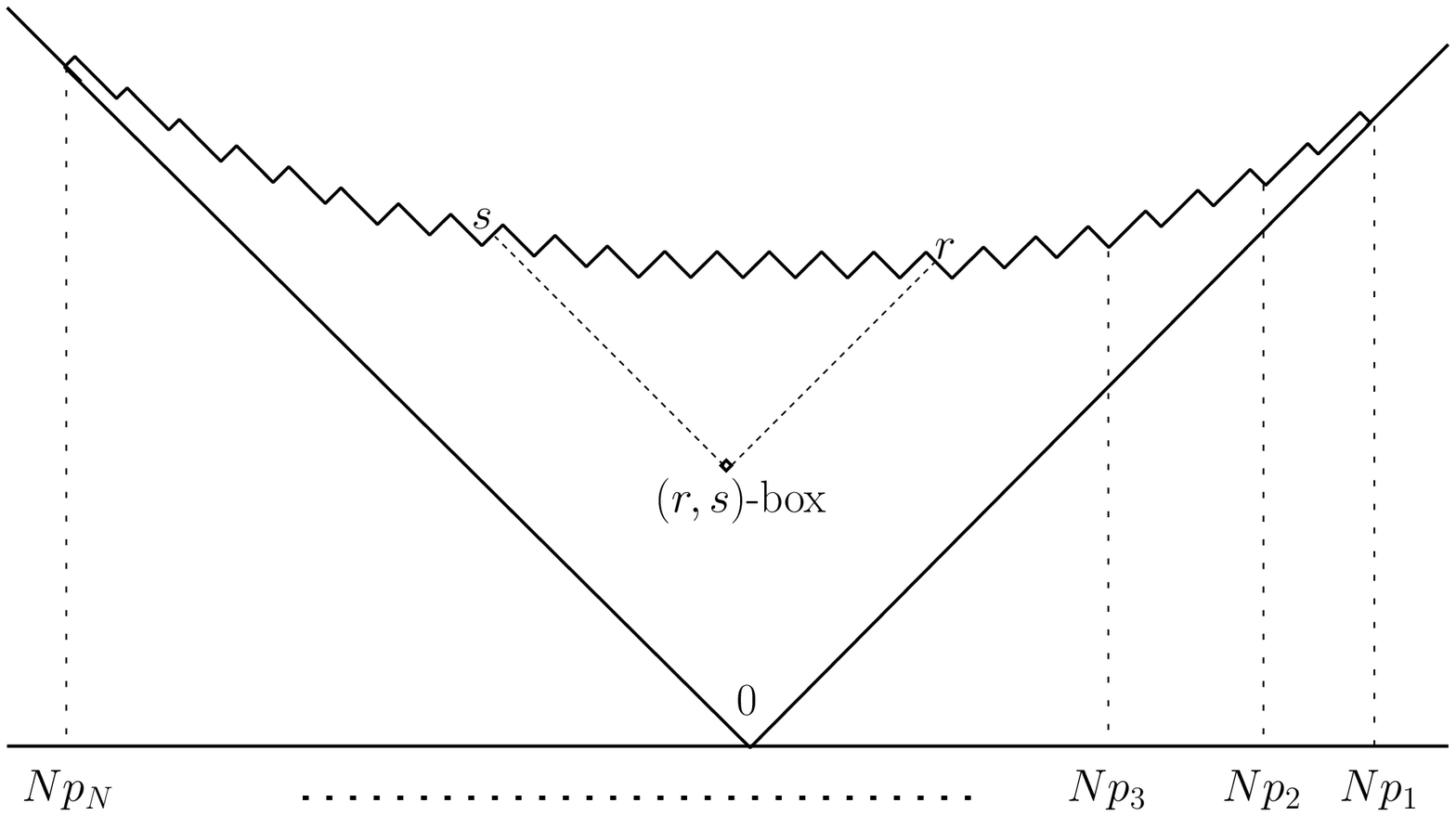}
\mycaption{8}{A general view of the $SU(N)$ ground partition.}
\end{center}
\end{figure}
We factor the products in (\ref{def Z SU(N) pert hook}) 
according to the color-coding of the partition as 
in the case of $N=3$. 
\begin{eqnarray}
\prod_{(i,j)\in\mu(\{\emptyset\};\{p_s\})}
\left(q^{-\frac{h(i,j)}{2}}
-q^{\frac{h(i,j)}{2}}\right)^{-1}
=\prod_{1\leq r<s\leq N} \ 
\prod_{(r,s)\textrm{-boxes}}
\left(q^{-\frac{h((r,s)\textrm{-box})}{2}}
-q^{\frac{h((r,s)\textrm{-box})}{2}}\right)^{-1}. 
\label{component N grand hook}
\end{eqnarray}
The contribution of the $(r,s)$-boxes 
can be computed in a form similar 
to (\ref{component N=3 grand hook}). 
We thus obtain the following expression 
for $\mathcal{Z}_q^{SU(N)\,pert}$;  
\begin{eqnarray}
&&
\mathcal{Z}_q^{SU(N)\,pert}(\{p_r\},Q)
\nonumber \\
&&~~~~~
=
\prod_{1\leq r<s\leq N}
\prod_{k=1}^{p_r-p_s-1}
\left\{
\frac{Q^{\frac{1}{2}}}
     {\Bigl(\prod_{t=r}^{s-1}Q_{F_t}\Bigr)^{-\frac{1}{2}}
       q^{\frac{Nk}{2}}
      -\Bigl(\prod_{t=r}^{s-1}Q_{F_t}\Bigr)^{\frac{1}{2}}
      q^{-\frac{Nk}{2}}}
\right\}^k,
\label{Z SU(N) pert}
\end{eqnarray}
where we put 
$Q_{F_r} \equiv 
q^{N(\widetilde{p}_r-\widetilde{p}_{r+1})}$ 
for $1 \leq r \leq N$.

%%%%%%%%%%%%%%%%%%%%%%%%%%%%%%%%%%%%%%%%%%%%%%%%%%%%%%%
\subsubsection*{$SU(N)$ instantons}

We factorize the partition function 
(\ref{Z U(1) N-component}) in the following form.    
\begin{eqnarray}
&&
Z_q^{U(1)}(Q)
\nonumber \\
&&~~
= 
\sum_{\{p_r\}}
q^{-\frac{N^2}{6}
\sum_{s=1}^{N}\widetilde{p}_s^3}  \, 
\mathcal{Z}_q^{SU(N)\,pert}(\{p_r\},Q)^2 \,
\sum_{\{\lambda^{(r)}\}}
\mathcal{Z}_q^{SU(N)\,inst}
(\{\lambda^{(r)}\},\{p_r\},Q)^2, 
\label{Z SU(N) factorization}
\end{eqnarray}
where 
$\mathcal{Z}_q^{SU(N)\,inst}$ 
are defined by the relations 
\begin{eqnarray}
&& 
Q^{\frac{1}{2}
   |\mu (\{\lambda^{(r)}\};\,\{p_r\})|}
s_{\mu(\{\lambda^{(r)}\};\,\{p_r\})}(q^{-\rho})
\nonumber \\
&&~~~~
=
q^{-\frac{N^2}{12}
\sum_{s=1}^{N}\widetilde{p_s}^3}\,
\mathcal{Z}_q^{SU(N)\,pert}(\{p_r\},Q)\,
\mathcal{Z}_q^{SU(N)\,inst}(\{\lambda^{(r)}\},\{p_r\},Q).
\label{def Z SU(N) inst}
\end{eqnarray}
Owing to (\ref{def Z SU(N) pert}) 
we can write the above as 
\begin{eqnarray}
\mathcal{Z}_q^{SU(N)\,inst}(\{\lambda^{(r)}\},\{p_r\},Q)
&=&
Q^{
\frac{1}{2}
\left(
    |\mu (\{\lambda^{(r)}\};\,\{p_r\})|
    -|\mu(\{\emptyset\};\,\{p_r\})|
\right) }
\frac{
      s_{\mu (\{\lambda^{(r)}\};\,\{p_r\})}
      (q^{-\rho})} 
     {s_{\mu (\{\emptyset \};\,\{p_r\})}
      (q^{-\rho})}. 
\label{def Z SU(N) inst schur}
\end{eqnarray}

Let us compute $\mathcal{Z}_q^{SU(N)\,inst}$. 
We order $p_1 > p_2 > \cdots >p_N$ for simplicity. 
The ratio of the Schur functions in 
(\ref{def Z SU(N) inst schur}) is translated to 
infinite products by applying the formula 
(\ref{product formula Schur}). 
By using the description (\ref{SU(N) Maya diagrams}) 
we can classify the ingredients of the products  
according as they are made of 
$q^{-\frac{N}{2}
   \left(x_i(\lambda^{(r)})-x_j(\lambda^{(r)})\right)} 
-q^{\frac{N}{2}
   \left(x_i(\lambda^{(r)})-x_j(\lambda^{(r)})\right)}$
where $1 \leq r \leq N$, 
or 
$(\prod_{t=r}^{s-1}Q_{F_t})^{-\frac{1}{2}}
q^{-\frac{N}{2}
    \left(x_i(\lambda^{(r)})-x_j(\lambda^{(s)})\right)}
-(\prod_{t=r}^{s-1}Q_{F_t})^{\frac{1}{2}}
q^{\frac{N}{2}
    \left(x_i(\lambda^{(r)})-x_j(\lambda^{(s)})\right)}$
where $1 \leq r < s \leq N$.
We then factor the RHS of (\ref{def Z SU(N) inst schur}) 
by this classification. 
Taking (\ref{N-1 size k}) into account, we obtain 
the following expression for $\mathcal{Z}_q^{SU(N)\,inst}$.
\begin{eqnarray}
&&
\mathcal{Z}_q^{SU(N)\,inst}(\{\lambda^{(r)}\},\{p_r\},Q) 
\nonumber \\*
&&
=
\pm Q^{\frac{N}{2}\sum_{r=1}^N |\lambda^{(r)}|}
q^{-\frac{N^2}{4}\sum_{r=1}^{N}\kappa(\lambda^{(r)})}
\prod_{r<s}^N
\Bigl(\prod_{t=r}^{s-1}Q_{F_t}\Bigr)
^{-\frac{|\lambda^{(r)}|-|\lambda^{(s)}|}{2}}
\nonumber \\*
&&
\times 
\prod_{r=1}^N \,
\prod_{1\leq i<j<\infty}
\left\{
\frac{
  q^{-\frac{N}{2}
        \left(
           x_i(\lambda^{(r)})-x_j(\lambda^{(r)})
        \right)}
  -q^{\frac{N}{2}
         \left( 
           x_i(\lambda^{(r)})-x_j(\lambda^{(r)})
        \right)}}
{q^{-\frac{N(j-i)}{2}}-q^{\frac{N(j-i)}{2}}}
\right\} 
\nonumber \\*
&& 
\times 
\prod_{r<s}^N \,
\prod_{1\leq i,j<\infty}
\left\{
\frac{
 \Bigl(\prod_{t=r}^{s-1}Q_{F_t}\Bigr)^{-\frac{1}{2}}
 q^{-\frac{N}{2}
    \left(x_i(\lambda^{(r)})-x_j(\lambda^{(s)})\right)}
 -\Bigl(\prod_{t=r}^{s-1}Q_{F_t}\Bigr)^{\frac{1}{2}}
 q^{\frac{N}{2}
    \left(x_i(\lambda^{(r)})-x_j(\lambda^{(s)})\right)}}
{\Bigl(\prod_{t=r}^{s-1}Q_{F_t}\Bigr)^{-\frac{1}{2}}
q^{-\frac{N(j-i)}{2}}
-\Bigl(\prod_{t=r}^{s-1}Q_{F_t}\Bigr)^{\frac{1}{2}}
q^{\frac{N(j-i)}{2}}}
\right\}  \,.
\nonumber \\
\label{Z SU(N) inst for gauge}
\end{eqnarray}

The expression (\ref{Z SU(N) inst for gauge}) 
makes it possible to rewrite $\mathcal{Z}_q^{SU(N)\,inst}$ 
in a form convenient to compare with the topological 
string amplitude. 
We first notice the identity   
\begin{eqnarray}
\prod_{1\leq i<j<\infty}
\left\{
\frac{
  q^{-\frac{N}{2}
        \left(
           x_i(\lambda^{(r)})-x_j(\lambda^{(r)})
        \right)}
  -q^{\frac{N}{2}
         \left( 
           x_i(\lambda^{(r)})-x_j(\lambda^{(r)})
        \right)}}
{q^{-\frac{N(j-i)}{2}}-q^{\frac{N(j-i)}{2}}}
\right\} 
&=&
(-)^{|\lambda^{(r)}|}
q^{-\frac{N}{4}\kappa(\lambda^{(r)})}
s_{\lambda^{(r)}}(q^{N\rho})\ ,
\nonumber \\ 
\label{SU(N) formula 1}
\end{eqnarray}
which follows from the product formula 
(\ref{product formula Schur}). 
The identity (\ref{SU(2) infinite product 2}) 
of the $SU(2)$ theory has the following generalization. 
\begin{eqnarray}
&&
\prod_{r<s}^N \,
\prod_{1\leq i,j<\infty}
\left\{
\frac{
 \Bigl(\prod_{t=r}^{s-1}Q_{F_t}\Bigr)^{-\frac{1}{2}}
 q^{-\frac{N}{2}
    \left(x_i(\lambda^{(r)})-x_j(\lambda^{(s)})\right)}
 -\Bigl(\prod_{t=r}^{s-1}Q_{F_t}\Bigr)^{\frac{1}{2}}
 q^{\frac{N}{2}
    \left(x_i(\lambda^{(r)})-x_j(\lambda^{(s)})\right)}}
{\Bigl(\prod_{t=r}^{s-1}Q_{F_t}\Bigr)^{-\frac{1}{2}}
q^{-\frac{N(j-i)}{2}}
-\Bigr(\prod_{t=r}^{s-1}Q_{F_t}\Bigr)^{\frac{1}{2}}
q^{\frac{N(j-i)}{2}}}
\right\} 
\nonumber \\
&&=
\prod_{r<s}^N
\left\{
q^{\frac{N}{4} 
   \left( \kappa(\lambda^{(r)})-\kappa(\lambda^{(s)})
   \right)}
\Bigl(
\prod_{t=r}^{s-1}Q_{F_t}
\Bigr)^{\frac{1}{2} 
    \left( |\lambda^{(r)}|+|\lambda^{(s)}|
    \right)}
\prod_{k=1}^{\infty}
\Bigl(1-
\Bigl(\prod_{t=r}^{s-1}Q_{F_t}\Bigr)q^{Nk}
\Bigr)^k
\right\} 
\nonumber \\
&&~~~
\times
\sum_{\nu^{(1)},\cdots,\,\nu^{(N-1)}}
\Bigl(
\prod_{t=1}^{N-1}
Q_{F_t}^{|\nu^{(t)}|}
\Bigr)
s_{\nu^{(1)}}(q^{N(\lambda^{(1)}+\rho)})
s_{\nu^{(N-1)}}
(q^{N(\widetilde{\lambda^{(N)}}+\rho)})
\nonumber \\
&&~~~~~~~~~~~~
\times 
\sum_{\chi^{(1)},\cdots,\,\chi^{(N-2)}}
\prod_{t=1}^{N-2}
s_{\nu^{(t)}/\chi^{(t)}}
(q^{N(\widetilde{\lambda^{(t)}}+\rho)})
s_{\nu^{(t+1)}/\chi^{(t)}}
(q^{N(\lambda^{(t)}+\rho)})\ .
\label{SU(N) formula 2}
\end{eqnarray}

The substitution of 
(\ref{SU(N) formula 1}) and (\ref{SU(N) formula 2}) into 
(\ref{Z SU(N) inst for gauge}) gives rise to some simple 
powers of $q$ and $Q_{F_r}$ in addition to a bunch  
of the Schur functions. The exponent of $q$ can be read as 
\begin{eqnarray}
&&
q^{-\frac{N^2}{4}\sum_{r=1}^N \kappa(\lambda^{(r)})}
\cdot 
q^{-\frac{N}{4}\sum_{r=1}^N \kappa(\lambda^{(r)})}
\cdot 
\prod_{r<s}^N
q^{\frac{N}{4}
\left(\kappa(\lambda^{(r)})-\kappa(\lambda^{(s)})\right)}
=
q^{-\frac{N}{2}
\sum_{r=1}^N r\kappa(\lambda^{(r)})}, 
\label{SU(N) exponent q}
\end{eqnarray}
while the exponents of $Q_{F_r}$ become 
\begin{eqnarray}
\prod_{r<s}^N
\Bigl(
\prod_{t=r}^{s-1}Q_{F_t}
\Bigr)^{|\lambda^{(s)}|}
=
\prod_{r=1}^{N-1} 
Q_{F_r}^{r\sum_{s=r+1}^N|\lambda^{(s)}|}.
\label{SU(N) exponent Qfr}
\end{eqnarray}
Together with the contributions 
of the Schur functions from (\ref{SU(N) formula 1}) 
and (\ref{SU(N) formula 2}) 
we finally obtain  
\begin{eqnarray}
&&
\mathcal{Z}_q^{SU(N)\,inst}(\{\lambda^{(s)}\},\{p_s\},Q) 
\notag\\*
&&
=
\pm Q^{\frac{N}{2}\sum_{r=1}^N |\lambda^{(r)}|}
q^{-\frac{N}{2}\sum_{r=1}^{N}r\kappa(\lambda^{(r)})}
\,\prod_{r=1}^{N-1} 
Q_{F_r}^{r\sum_{s=r+1}^N |\lambda^{(s)}|}
\,\prod_{r=1}^N 
s_{\lambda^{(r)}}(q^{N \rho})
\nonumber \\
&&~~~
\times
\prod_{r<s}^N 
\prod_{k=1}^{\infty}
\Bigl(1-
\Bigl(\prod_{t=r}^{s-1}Q_{F_t}\Bigr)q^{Nk}
\Bigr)^k 
\nonumber \\
&&~~~ 
\times 
\sum_{\nu^{(1)},\cdots,\,\nu^{(N-1)}}
\Bigl(
\prod_{t=1}^{N-1}Q_{F_t}^{|\nu^{(t)}|}
\Bigr)
s_{\nu^{(1)}}
(q^{N\left(\lambda^{(1)}+\rho\right)})
s_{\nu^{(N-1)}}
(q^{N(\widetilde{\lambda^{(N)}}+\rho)})
\nonumber \\
&&~~~~~~~~~~~~~~~
\times 
\sum_{\chi^{(1)},\cdots,\,\chi^{(N-2)}}
\prod_{t=1}^{N-2}
s_{\nu^{(t)}/\chi^{(t)}}
(q^{N(\widetilde{\lambda^{(t)}}+\rho)})
s_{\nu^{(t+1)}/\chi^{(t)}}
(q^{N(\lambda^{(t)}+\rho)}).
\label{Z SU(N) inst for string}
\end{eqnarray}

%%%%%%%%%%%%%%%%%%%%%%%%%%%%%%%%%%%%%%%%%%%%%%%%%%
\subsubsection*{Interpretation as five-dimensional 
$SU(N)$ Yang-Mills}

We fix the $U(1)$-charges $p_r$.  
The relevant field theory parameters are 
$a_r,\Lambda$ and $R$,   
where $a_r$ $(r=1,\cdots,N)$  
are the VEVs of the adjoint scalar 
in the vector multiplet.
We identify $q, Q_{F_r}$ and $Q$ with 
$a_r,\Lambda$ and $R$ as follows. 
\begin{eqnarray}
q=e^{-\frac{2}{N}R\hbar},~~~
Q_{F_r}=e^{-2R(a_r-a_{r+1})},~~~
Q=-(2R\Lambda)^2. 
\label{SU(N) FT parameter}
\end{eqnarray}
Since we have set 
$Q_{F_r}=q^{N(\widetilde{p}_r-\widetilde{p}_{r+1})}$ 
the above implies 
\begin{eqnarray}
a_r=\hbar \widetilde{p}_r.~~~~~ 
\label{SU(N) a and p}
\end{eqnarray}

By the identifications (\ref{SU(N) FT parameter}),  
$\mathcal{Z}_q^{SU(N)\,inst}$ are translated 
to the instanton contributions in five-dimensional 
gauge theories. 
This follows by rephrasing 
the expression (\ref{Z SU(N) inst for gauge})
in terms of the field theory parameters. 
In particular, we obtain  
\begin{eqnarray}
\sum_{\{\lambda^{(r)}\}}
\mathcal{Z}_q^{SU(N)\,inst}
(\{\lambda^{(r)}\},\{p_r\},\,Q)^2
&=&
Z^{SU(N)\,inst}_{\,5d\,\mbox{\tiny{SYM}}}
(\{a_r\};\,\Lambda,R,\hbar).  
\label{Z SU(N) SYM inst}
\end{eqnarray}
The RHS is the instanton part of the partition 
function \cite{Nekrasov-Okounkov} for five-dimensional 
${\cal N}=1$ supersymmetric $SU(N)$ Yang-Mills 
with the Chern-Simons term \cite{Chern-Simons}. 
The Chern-Simons corrections come from 
the following factor in (\ref{Z SU(N) inst for gauge}). 
\begin{eqnarray}
q^{-\frac{N^2}{4}\sum_{r=1}^{N}\kappa(\lambda^{(r)})}
\prod_{1\leq r<s \leq N}
\Bigl(\prod_{t=r}^{s-1}Q_{F_t}\Bigr)
^{-\frac{|\lambda^{(r)}|-|\lambda^{(s)}|}{2}}, 
\label{SU(N) CS correction}
\end{eqnarray}
which is also understood \cite{Chern-Simons framing}
as a part of the so-called framing factor 
of the topological string vertices 
\cite{topological vertex}. 
The Chern-Simons coupling constant becomes 
$c_{cs}=N$.

Thanks to the identifications (\ref{SU(N) FT parameter}),  
the square of $\mathcal{Z}_{q}^{SU(N)\,pert}$ takes 
a form analogous to (\ref{ln Z by g}).  
By using the expression (\ref{Z SU(N) pert}) it becomes 
\begin{eqnarray}
\mathcal{Z}_{q}^{SU(N)\,pert}(\{p_{r}\},Q)^{2}
&=&
\prod_{r<s}^N
\exp
\biggl\{
-g(p_{r}-p_{s}|\hbar,2R,\Lambda)
-g(p_{r}-p_{s}|-\hbar,2R,\Lambda)
\biggr\}. 
\label{Z SU(N) pert for gauge}
\end{eqnarray}
As is the case of $SU(2)$, the functions 
$g(p_{r}-p_{s}|\hbar,2R,\Lambda)$ coincide with 
$\gamma_{\hbar}
(\hbar(\tilde{p}_{r}-\tilde{p}_{s})|2R,\Lambda)$ 
up to linear functions 
in $\hbar(\tilde{p}_{r}-\tilde{p}_{s})$. 
Thus, 
by taking (\ref{SU(N) a and p}) into account, 
we obtain 
\begin{eqnarray}
\mathcal{Z}_{q}^{SU(N)\,pert}(\{p_{r}\},Q)^2
&=&
\exp
\Biggl\{
-\Delta_{N}^{(0)}(\hbar,2R,\Lambda)
-\sum_{\scriptstyle r\neq s}^N
\Delta_{N\,r,s}^{(1)}
(\hbar,2R,\Lambda)
(a_{r}-a_{s})
\Biggr\}
\notag\\
&&
\times
\exp
\Biggl\{
-\sum_{\scriptstyle r\neq s}^N
\gamma_{\hbar}(a_{r}-a_{s}|2R,\Lambda)
\Biggr\}.
\label{Z SU(N) pert for gauge 2}
\end{eqnarray}
The expression (\ref{Z SU(N) pert for gauge 2}) 
shows that the square of $\mathcal{Z}_{q}^{SU(N)\,pert}$
is identified with the perturbative part of the partition 
function for five-dimensional supersymmetric $SU(N)$ Yang-Mills. 
Let us recall that the perturbative part is introduced 
in (\ref{def Z SU(N) pert}) as 
the Boltzmann weight for the ground partition 
after removing the factor,  
$q^{-\frac{N^2}{12}\sum_{r=1}^N\widetilde{p}_r^3}$. 
This removed factor turns to be the perturbative 
correction \cite{triple intersection} from the Chern-Simons term. 
Together with (\ref{Z SU(N) SYM inst}) we finally find out   
\begin{eqnarray}
&&
q^{-\frac{N^2}{6}\sum_{r=1}^N\widetilde{p}_r^3}
\mathcal{Z}_q^{SU(N)\,pert}(\{p_r\},Q)^2
\sum_{\{\lambda^{(r)}\}}
\mathcal{Z}_q^{SU(N)\,inst}(\{\lambda^{(r)}\}, 
\{p_r\},Q)^2
\nonumber \\*[1.5mm]
&&
~~~~~~~~~~~~~~~~~~~~~~~~~~~~~~~~~~~~~~~~~~~~~~~
=
Z^{SU(N)}_{\,5d \, \mbox{\tiny{SYM}}}
(\{a_r\};\,\Lambda,R,\hbar). 
\label{Z SU(N) SYM}
\end{eqnarray}
The RHS is the exact partition function 
\cite{Nekrasov-Okounkov} for the five-dimensional supersymmetric  
$SU(N)$ Yang-Mills with the Chern-Simons term.

Similarly to the $SU(2)$ case, 
the gauge theory is realized as the $\hbar \rightarrow 0$ 
limit of the above partition function.  
It is the thermodynamic limit with $a_r, \Lambda$ and $R$ fixed. 
The corresponding ground partition in 
(\ref{Z SU(N) SYM}) becomes very large and 
the $U(1)$ instantons $\lambda^{(r)}$ 
provide the deformation of the ground partition. 
Hence we can consistently say that 
the ground partitions describe the perturbative regime 
of the Coloumb branch while their deformation is non-perturbative 
in the gauge theory.

%%%%%%%%%%%%%%%%%%%%%%%%%%%%%%%%%%%%%%%%%%%%%%%%%
\subsubsection*{Ground partitions and classical $SU(N)$ 
geometries}

The relevant non-compact Calabi-Yau threefold is 
an ALE space with $A_{N-1}$ singularity fibred over 
$\mathbb{P}^1$. 
Type of the fibrations is labelled by 
the framing $m \in [0,\,N]$. 
We choose the framing to be $m=N$. 
The geometrical data are the K\"ahler parameters 
$t_B$ and $t_{F_r}$, where $1 \leq r \leq N-1$. 
They correspond respectively to 
the K\"ahler volumes of 
the base $\mathbb{P}^1$ and the $N-1$ blow-up cycles 
of $A_{N-1}$ singularity in the fibre.

The topological string amplitude on this local geometry 
is given in (\ref{SU(N) string amplitude}). 
We denote the amplitude as 
$Z^{SU(N)}_{string}(q,\{Q_{F_r}\},Q_B)$, where 
$q=e^{-g_{st}},Q_{F_r}=e^{-t_{F_r}}$ and 
$Q_B=e^{-t_B}$.  
By comparing (\ref{Z SU(N) inst for string}) with 
(\ref{SU(N) string amplitude}), we see that 
the gauge instanton contribution 
(\ref{Z SU(N) SYM inst}) is converted 
to the string amplitude as follows.
\begin{eqnarray}
&&
\sum_{\{\lambda^{(r)}\}}
\mathcal{Z}_q^{SU(N)\,inst}
(\{\lambda^{(r)}\},\{p_r\},Q )^2
\nonumber \\
&&~~~~
=\,
\prod_{r<s}^N
\prod_{k=1}^{\infty}
\Bigl(1-
   \Bigl(\prod_{t=r}^{s-1}Q_{F_t}\Bigr) 
   q^{Nk}
\Bigr)^{2k}
~ 
Z_{string}^{SU(N)}
(q^N,\{Q_{F_r}\},Q^N), 
\label{SU(N) ins vs string}
\end{eqnarray}
where we put 
$Q_{F_r}=q^{N(\widetilde{p}_r-\widetilde{p}_{r+1})}$.  
This is in accord with the result 
\cite{Iqbal,Eguchi} that 
the topological string amplitude 
on this local Calabi-Yau geometry leads to 
the instanton part of the partition function 
of five-dimensional supersymmetric $SU(N)$ Yang-Mills.

Due to the identification $\hbar \sim g_{st}$,  
the thermodynamic limit is in the perturbative regime 
of string theory. 
The ground partitions at the thermodynamic limit 
should be interpreted as classical objects in string theory. 
The relations $t_{F_r}=2R(a_r-a_{r+1})$, 
which follow from (\ref{SU(N) ins vs string}), 
suggest that the $SU(N)$ ground partitions describe 
the resolutions of $A_{N-1}$ singularity 
in this local geometry. 
This identification allows us to interpret 
the perturbative Chern-Simons correction 
as the triple intersections of 
the four-cycles \cite{triple intersection}
in the local geometry.

%%%%%%%%%%%%%%%%%%%%%%%%%%%%%%%%%%%%%%%%%%
%%%%%%%%%%%%%%%% APPENDIX %%%%%%%%%%%%%%%%
%%%%%%%%%%%%%%%%%%%%%%%%%%%%%%%%%%%%%%%%%%
\appendix
\section{Topological string amplitudes}
\label{appendix}

We present the topological string amplitudes  
related to four-dimensional $\mathcal{N}=2$ 
supersymmetric gauge theories. 
The geometric engineering 
\cite{Geometric engineering} dictates that  
four-dimensional $\mathcal{N}=2$  
supersymmetric $SU(N)$ Yang-Mills is realized by 
an ALE space with $A_{N-1}$ singularity 
fibered over $\mathbb{P}^1$.
The fibration over $\mathbb{P}^1$ 
is labeled by an integer $m\in [0,N]$, 
which is often called the framing.
We consider the case of $m=N$.
The geometric data turn to be related with 
the field theory parameters.
The K\"ahler parameter $t_B$ of the base 
$\mathbb{P}^1$ is proportional 
to $1/g^2$, 
where $g$ is the gauge coupling constant 
at the string scale.  
The gauge instanton effect is weighted with 
$e^{-c_2/g^2}\sim \Lambda^{2Nc_2}$.
This leads to $e^{-t_B}\sim\Lambda^{2N}$.
The K\"ahler parameters $t_{F_r}$
of the blow-up cycles in the fibre 
are proportional to $a_r$, the VEVs of the adjoint 
scalar in the vector multiplet.

The topological vertex \cite{topological vertex} 
provides a powerful method to compute all genus 
topological $A$-model partition functions 
on local toric Calabi-Yau threefolds.
The computations are carried out 
by using diagrammatic techniques like the Feynman rules. 
The diagrams are the dual toric diagrams consisting of 
trivalent vertices. Topological string amplitudes 
on $\mathbb{C}^3$ with open-string boundary 
conditions are attached to the vertices. 
They are derived based on the large 
$N$ duality \cite{Large N} and 
the boundary conditions turn out to be fixed 
by three partitions. 
Gluing the vertices by a certain rule, 
one obtains the topological string amplitudes on 
the threefolds.

The diagram which describes the above local geometry 
for the $SU(N)$ gauge theory is depicted in Figure 9. 
The method of the topological vertex gives the amplitude 
in the following form
\footnote{Our convention is slightly different from 
\cite{Iqbal,Eguchi}.}. 
\begin{eqnarray}
&&
Z_{string}^{SU(N)}(q,\{Q_{F_r}\},Q_B)
\notag\\*[2mm]
&&
=\sum_{\lambda^{(1)},\cdots,\,\lambda^{(N)}}
\left[
\begin{array}{l}
Q_B^{\frac{1}{2}\sum_{r=1}^{N}|\lambda^{(r)}|}\, 
q^{-\frac{1}{2}\sum_{r=1}^{N} r\kappa(\lambda^{(r)})}\
{\displaystyle 
     \prod_{r=1}^{N-1} 
     Q_{F_r}^{r\sum_{s=r+1}^N|\lambda^{(s)}|}}\ 
{\displaystyle 
    \prod_{r=1}^{N} 
    s_{\lambda^{(r)}}(q^{\rho})}
\\*[1.5mm]
\times 
{\displaystyle
\sum_{\nu^{(1)},\cdots,\,\nu^{(N-1)}}
\Bigl( 
\prod_{t=1}^{N-1}Q_{F_t}^{|\nu^{(t)}|} 
\Bigr)}\
s_{\nu^{(1)}}
(q^{\lambda^{(1)}+\rho})
s_{\nu^{(N-1)}}
(q^{\widetilde{\lambda}^{(N)}+\rho})
\\*[2.0mm]
%\\*[1.5mm]
\ \ \ \ \ \ 
\times 
{\displaystyle  
\sum_{\chi^{(1)},\cdots,\,\chi^{(N-2)}}
\prod_{t=1}^{N-2}}\
s_{\nu^{(t)}/\chi^{(t)}}
(q^{\widetilde{\lambda^{(t)}}+\rho})
s_{\nu^{(t+1)}/\chi^{(t)}}
(q^{\lambda^{(t)}+\rho})
\label{SU(N) string amplitude}
\end{array}
\right]^2,
\end{eqnarray}
where the parameters $q,Q_{F_r}$ and $Q_B$ are given by 
\begin{eqnarray}
q=e^{-g_{st}},~~~
Q_{F_r}=e^{-t_{F_r}},~~~
Q_{B}=e^{-t_B}. 
\label{parameters of string}
\end{eqnarray}

(\ref{SU(N) string amplitude}) 
for the case of $SU(2)$ becomes as follows; 
($Q_F= Q_{F_1}$)
\begin{eqnarray}
&&
Z_{string}^{SU(2)}(q,Q_{F},Q_B)
\notag\\*[2mm]
&&
=\sum_{\lambda^{(1)},\lambda^{(2)}}
\left[
\begin{array}{l}
Q_B^{\frac{1}{2} 
   \left(|\lambda^{(1)}|
           +|\lambda^{(1)}| \right)}\, 
q^{-\frac{1}{2}
   \left(\kappa(\lambda^{(1)})
          +2\kappa(\lambda^{(2)})\right)}
\  Q_{F}^{|\lambda^{(2)}|}
\\[1.5mm]
\ \  
\times 
s_{\lambda^{(1)}}(q^{\rho}) \, 
s_{\lambda^{(2)}}(q^{\rho}) \, \,
{\displaystyle\sum_{\nu}Q_{F}^{|\nu|}}
\,
s_{\nu}(q^{\lambda^{(1)}+\rho})
s_{\nu}(q^{\widetilde{\lambda^{(2)}}+\rho})
\end{array}
\right]^2. 
\label{SU(2) string amplitude}
\end{eqnarray}
\begin{figure}[t]
\begin{center}
\includegraphics[scale=0.8]{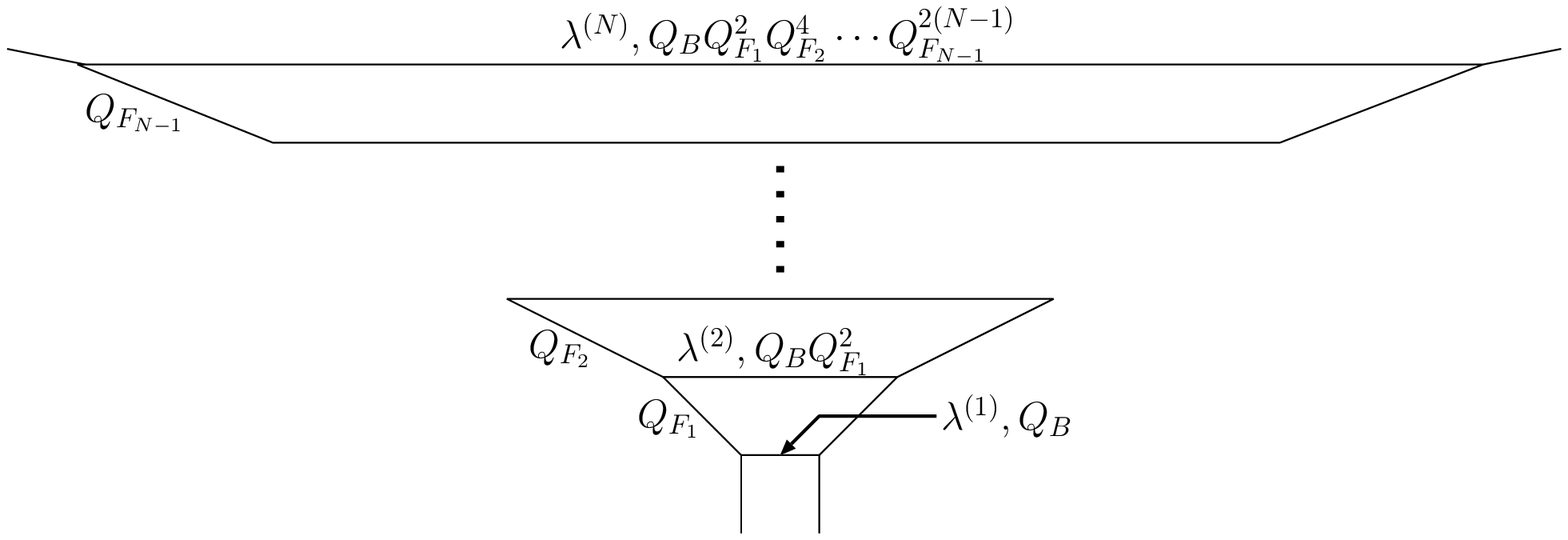}
\mycaption{9}{The diagram for the local geometry of 
the SU(N) gauge theory. $Q_{i}=e^{-t_i}$, 
where $t_i=t_B,t_{F_r}$ ($1 \leq r \leq N-1$).
Partitions $\lambda^{(r)}$ are attached to 
the corresponding $\mathbb{P}^1$s.}
\end{center}\label{SU(N) geometry}
\end{figure}
%
%

%%%%%%%%%%%%%%% ACKNOWLEDGEMENTS %%%%%%%%%%%%%%%
\subsection*{Acknowledgements}
T.N. is supported in part by Grant-in-Aid for 
Scientific Research 15540273. 
K.T. is supported in part by Grant-in-Aid for 
Scientific Research 16340040.

%%%%%%%%%%%%%%%%%%%%%%%%%%%%%%%%%%%%%%%%%%
%%%%%%%%%%%%%%% REFERENCES %%%%%%%%%%%%%%%
%%%%%%%%%%%%%%%%%%%%%%%%%%%%%%%%%%%%%%%%%%

%%%%%%%%%%%%%%%%%%%%%%%%%%%%%%%%%%%%%%%%%%%%
%%%%%%%%%%%%%%% END DOCUMENT %%%%%%%%%%%%%%%
%%%%%%%%%%%%%%%%%%%%%%%%%%%%%%%%%%%%%%%%%%%%
\end{document}